\documentclass[structabstract]{aa}  
\usepackage{graphicx}
\usepackage{txfonts}
\usepackage{lscape}
\usepackage{subfigure}
\begin{document}
   \title{Long-term variations in the mean meridional motion of the\break sunspot groups}


   \author{J. Javaraiah}

   \institute{Indian Institute of Astrophysics, Bangalore- 560034, India\\
              \email{jj@iiap.res.in}
             }

   \date{Received ; accepted }

 
  \abstract
   {}
{We seek 
  the long-term variations close to the length 
of a solar cycle  in the mean meridional motion of 
sunspot groups (a proxy of the   
meridional plasma flow).}
   {Using the largest set of available reliable sunspot group data, 
the combined  Greenwich and
 Solar Optical Observation Network sunspot group
data during the period 1879\,--\,2008, 
we determined  variations in the mean meridional motion of
the sunspot groups in 
the  Sun's whole northern and southern 
hemispheres and also in different $10^\circ$ latitude intervals.  
We determined the variations from   the 
yearly  data 
and  for the sake of better statistics  by 
 binning the  
data into 
 3\,--\,4 year moving time intervals (MTIs)  successively shifted  
by one year.
We  determined the periodicities in the mean meridional  motion from  
the  fast  
Fourier transform (FFT) power spectrum analysis.   The values of the
periodicities are determined 
 from the maximum entropy method (MEM) and   
 the temporal dependencies of the periodicities are determined from   
 the Morlet-wavelet analyses.} 
{We find that the mean  meridional motion of the spot
groups varies considerably on a time  scale  of  about 5\,--\,20 years.
The maximum  
 amplitude of the variation 
is about  10\,--\,15 m s$^{-1}$   in both the northern and the southern
 hemispheres.
Variation in the mean motion is considerably 
different during different solar 
cycles. 
 At the maximum epoch (year 2000) of the 
current cycle~23, the mean  motion
is relatively strong in the past  100 years  
and northbound in both the
 northern and the southern hemispheres. This abnormal behavior of the  
  mean  motion   may be related to 
 the low strength and the long duration of the current cycle, and also to  
 the  violation of
the Gnevyshev
and Ohl rule by the cycles pair 22,23.
The power spectral analyses suggest the existence of $\approx$ 
3.2- and $\approx$ 4.3-year
 periodicities in the mean  motion  
of the spot groups  in the southern hemisphere, whereas 
 a  13\,--\,16 year  periodicity is found to exist 
   in the mean  motion 
of the northern hemisphere.
There is  strong evidence for a latitude-time dependency in the  
 periodicities of the mean motion. 
 The north-south  difference
 in  the mean  motion  
 also varies by about 10 m s$^{-1}$.  During 
the recent cycles, the north-south difference is
negligibly  small. 
Approximate  12- and  
22-year periodicities  are  found to  exist 
 in the north-south difference.
 The implications of all 
these results 
are 
briefly discussed.}
    {}
\authorrunning{J. Javaraiah}               
\titlerunning{Variations in the mean meridional motion of sunspot groups}

   \maketitle
%

\section{Introduction}
The study of the  variations in the meridional flows 
is important  for understanding the underlying mechanism of 
 the solar cycle 
(Babcock 1961;  Ulrich \& Boyden 2005).
Surface Doppler measurements 
and helioseismology measurements of the surface and the subsurface flows
  (e.g., Ulrich \& Boyden 2005; 
 Gonz\'alez Hern\'andez et al. 2008)
 suggest  poleward flows of about 10\,--\,20 m s$^{-1}$
and a considerable north-south asymmetry in the meridional flow.  
Surface Doppler measurements are 
available  for about 3 cycles (Ulrich \& Boyden 2005).
Doppler measurements suffer from errors  caused by   the scatter light and 
the B-angle influences (Beckers 2007).  
The motions of many magnetic tracers, particularly sunspots,  
have been  used for a long time  as a proxy of the fluid motions to  
 study the solar rotational and  the meridional flows (Schr\"oter 1985; Javaraiah \& Gokhale 2002).
The sunspot    
data have been available for more than  
100 years. 
 However, the derived rates of rotation and  meridional flows  
  depend on the method of the selection of the spots or the spot groups   
(e.g., Howard et al. 1984;  Balthasar et al. 1986; 
 Zappal\'a \& Zuccarello 1991; Zuccarello 1993; 
Javaraiah \& Gokhale 1997a; Javaraiah 1999; Hiremath 2002; 
Sivaraman et al. 2003).  Proper motions and evolutionary factors 
of the spot groups may also   influence 
 the derived rates of the flows to some extent. The proper motions 
are random in nature, hence their effect can be reduced with 
the use of a large data 
set.  
   Recently, Ru\v{z}djak et al. (2005) have found 
that evolutionary factors of the sunspot groups in the determination of the 
positions of the spot groups is small in the estimated mean meridional motion
of the spot groups.  

A number of scientists studied the solar cycle variations of 
the mean meridional motion of the spot groups (see Javaraiah \& Ulrich 2006).
Since  yearly data are inadequate  for this purpose,
 particularly around a solar cycle 
minimum,
  some authors have used the
superposed epoch analysis of the data during a few or all   cycles
for which the data were available and studied  
the mean solar cycle variation
 (Balthasar et al. 1986;  Howard \& Gilman  1986).
 Recently, by using the same method,  Javaraiah \& Ulrich (2006)
 analyzed the combined  Greenwich and
 Solar Optical Observation Network sunspot group
data during the period 1879\,--\,2004 and 
determined the mean solar cycle variation in the
  meridional motions of the sunspot groups. 
In that early paper, the spot group data during the cycles
12\,--\,20 were superposed according to the
 minima of these cycles.
 This  yielded an average solar cycle
 variation of the meridional motion over about 5\,--\,9 cycles, which  
suggests that only around the end of a solar cycle
 the meridional motion is considerably significant 
from zero and it is poleward in both the northern and the 
southern hemispheres. 
However,
  during some individual cycles the variations may 
 be considerably different
from  this average solar cycle variation.
 Javaraiah \& Ulrich (2006) also 
 determined
the cycle-to-cycle
variation of the mean (over the duration of a cycle)  meridional
motion of the spot groups during cycles 12\,--\,23 and  found  the 
 existence of  a weak long-period cycle (Gleissberg cycle) 
 in the cycle-to-cycle variation of the mean motion.
Since the meridional speed is  
negligibly small or zero in some phases  of a cycle or even with 
opposite signs during different phases of some  cycles,
  hence,  the motions are washed out in the average over the
cycle. 
 In order to get rid of this problem in the study of long-term variations 
in the mean meridional motion, it is necessary to  
 analyse   
 the spot or the spot group data  in the intervals considerably    
shorter than the length of a solar cycle. 

In
 the present paper  we have analyzed  the annual 
 spot group data during 1879\,--\,2008 and determined the 
variations in the mean meridional motions of the spot 
groups in the northern and the southern hemispheres.  
 As expected the statistics is  
 poor in case of the results derived from the annual data, particularly
at the  cycles minima. 
We  have taken some additional precautions, so  
 it is possible  to see the patterns of  
 the variations when they are close to  a solar cycle length, 
even in the annual time
series. 
However, 
we have also determined the  variation in the  mean
meridional motion of the spot groups by binning the spot group data into   
 the moving-time intervals (MTIs) of lengths (3\,--\,4 years) considerably
 greater 
than a year, but
 reasonably smaller 
than the length of a cycle; $i.e.$, less than the half of the
length of a cycle. In such a time-series which comprised the 
longer time intervals, it is relatively easy to detect 
the long-term variations near the length of a solar cycle
 (Javaraiah \& Gokhale 1995, 1997b). In addition, 
  the sizes of these series 
 reasonably large.  Hence, it enabled us to find  
the  periodicities that approach  
the length of a solar cycle
from the power spectral 
analysis.

In the next section we  describe the data and analysis. 
In Sect.~3 we present the variations in the 
 mean meridional motion of the  
spot groups  in the  whole northern  and 
the southern hemispheres,   
 as well as in different $10^\circ$ latitude intervals--and 
the corresponding differences between the whole northern and the southern 
hemispheres--during the period 1879\,--\,2008 and point out their
 important features.
In the same section, 
 we show  the periodicities
in the mean meridional motion of the spot groups 
 from the traditional FFT analyses.   From the    
MEM analyses, we  determined the values of the
 periodicities, and from the Morlet-wavelet analyses 
 we  determined the temporal dependencies of 
the detected periodicities.
In Sect.~4 we  summarize the results and the conclusions, 
and briefly  discuss
  the implications of these  results for understanding the solar 
 long-term variability. 

\vspace{0.3cm}
\section{Data and analysis}
  We have used the combined  Greenwich  
(1879\,--\,1976) and  Solar Optical Observation 
Network (SOON) (1977\,--\,2008) sunspot group
data, which are    
 taken  from  David Hathaway's  website 
{\tt http://solarscience.msfc.nasa.gov/greenwch.shtml}.
These data  
include the observation time (the Greenwich data contain the date with the
fraction of the day,  the SOON data do not contain 
 the fraction of 
the day), 
heliographic latitude 
and longitude, and central meridian distance (CMD), etc., of the spot 
groups for each day of observation.
The 
positions of the groups  are geometrical  positions of the 
centers of the groups.
The Greenwich data were compiled from the majority of the white
light photographs, which were secured at the Royal Greenwich Observatory
and at the Royal Observatory, Cape of Good Hope. The gaps in their
observations were filled with photographs from  other observatories,
 including the Kodaikanal Observatory, India.
The SOON data  included measurements made   by the 
United States Air Force (USAF) from 
the sunspot drawings of a network of the observatories
 that  includes telescopes
in Boulder, Colorado, Hawaii, etc.
 David Hathaway  scrutinized 
the Greenwich and SOON data and produced a reliable 
continuous data series
from 1874 up to now.  
 However, in this corrected  data  there may be some  minor 
differences mainly in the values of the  areas  of the spot groups of the 
two datasets   
(Hathaway \& Choudhary 2008).
 We did not use the data before 1879 because of the 
large uncertainties in the 1878 Greenwich data (Balthasar et. al. 1986;
Javaraiah \& Gokhale 1995).

 Here  the data reduction and the determination 
of the meridional motions 
of the spot groups is the same as  described in 
 Javaraiah \& Ulrich (2006). 
The data on the  recurrent and the non-recurrent spot groups are combined. 
The meridional velocity (daily rate of the latitudinal drift) of a spot 
group is computed using the difference between the epochs of its observation 
in 
consecutive days and the  heliographic 
latitudes of the spot group at these epochs. 
 For the sake of convenience in determining the north-south difference in the 
mean meridional motion of the spot groups, we  
used  the sign convention,
 {\it a positive value represents 
the northbound 
motion in both the northern and the southern hemispheres}.
As in  the paper referred to above,   
we have taken the following precaution to  
 substantially reduce the
 uncertainties in the derived results (Ward 1966;
Javaraiah \& Gokhale 1995):
We  excluded the
data corresponding to the $|CMD| > 75^\circ$ on any day of the spot group's
life span.
Furthermore, we  excluded the data corresponding to
 the `abnormal'  motions,  e.g.,  displacements
exceeding $3^\circ$ day$^{-1}$
  in the longitude or  $2^\circ$ day$^{-1}$ in the latitude. 
In addition, we did not use the data  corresponding to   the time-difference  
$>$ 2 days  of the spot group's life span. 

We  determined  the variations in the mean  
 meridional motion 
 of the sunspot groups  
in the  whole northern hemisphere and in the whole 
southern  hemisphere, 
 and also  in the separate  
$10^\circ$ latitude intervals.  
The variations in the mean motion of the spot groups  in a whole hemisphere
are determined from    the yearly data and  from the data in  3- and 4-year 
 MTIs during the period 1979\,--\,2008.  
For the separate $10^\circ$ latitude intervals, we   used only the  
4-year MTIs because, in a shorter than  4-year interval, 
 the data are found to be inadequate and the error bars are too  large 
 to plot the results, particularly during the cycle minima. 
   The accuracy in  determining the heliographic coordinates of
 the mass center
of a sunspot group can be estimated as close to $0.5^\circ$ and therefore
 error in
the calculations of daily velocity values can be estimated to 
 $\approx 1^\circ$ day$^{-1}$
($\approx 1.4 \times 10^4$ cm s$^{-1}$) (Balthasar \& W\"ohl 1980; 
Zappal\'a \& Zuccarello 1991). However, error in determining  the mean
velocity is inversely proportional to the square root of the number of
 observations
 used, and for example, it is on the order of $0.07^\circ$ day$^{-1}$
 ($9.9 \times 10^2$ cm s$^{-1}$) for 200  observations. 
The  number of data points are considerably more in several years of the 
annual series and in most of   the 
intervals  of the aforementioned  MTIs series. 
 As a result, the determined mean velocity is highly accurate  with 
respect to the mentioned error. 
 To determine the periodicities in the mean meridional motion 
of the spot groups using   the power spectral techniques, we  
applied some corrections to the original data,  described in the 
next section, where we describe the time series.

 Since the lengths of the  time series are inadequate for precisely  measuring 
the values of $\ge$ 11-year periodicities from the FFT analysis,    
 the uncertainties are large in the such 
 periodicities determined here  from the FFT analysis.
A  
 different approach to determining the values of the periodicities 
in a short time series with  higher accuracy is to compute
the power spectrum using MEM.
MEM analysis is a parametric modeling approach to  estimating the power 
spectrum of a time series. The method is data adaptive, since it is
 used upon an 
autoregressive modeling process. 
An important step in this method is the optimum selection 
of the order ($M$) of the autoregressive process,
 which is the number of immediately previous points that are used 
in calculating  a new point. 
If $M$ is chosen too low, the spectrum is over-smoothed and the high resolution 
potential is lost. If $M$ is chosen too high, frequency shifting and 
spontaneous splitting 
 of the spectral peaks occurs.  
The MEM code that we used here takes the values for $M$ 
 in the range ($n/3$, $n/2$) (Ulrych \& Bishop 1975)   or $2n/\ln(2n)$ 
(Berryman 1978),  
where $n$ is the total number of intervals in the 
analyzed time series. 
 To find the correct values of the periodicities 
found in the FFT power spectrum, 
 we  computed  
MEM power spectra by choosing  various 
values for   $M$ in the range ($n/3$, $n/2$) and  $2n/\ln(2n)$. 
We find that $M = n/3$ is  suitable in the present MEM 
analysis;  $i.e.$, in the derived spectra the peaks are considerably  sharp 
and well separated.

The wavelet analysis is a powerful method for 
analyzing localized variations 
in the power within a time series at many different 
frequencies (Torrence \& Compo 1998).
We did  the Morlet wavelet analysis to determine the temporal
 dependencies 
in  the periodicities found in the mean meridional motion 
of the spot groups from the FFT and MEM analysis.

\section{Results}

\begin{figure*}
\centering
\includegraphics[angle=90,width=\textwidth]{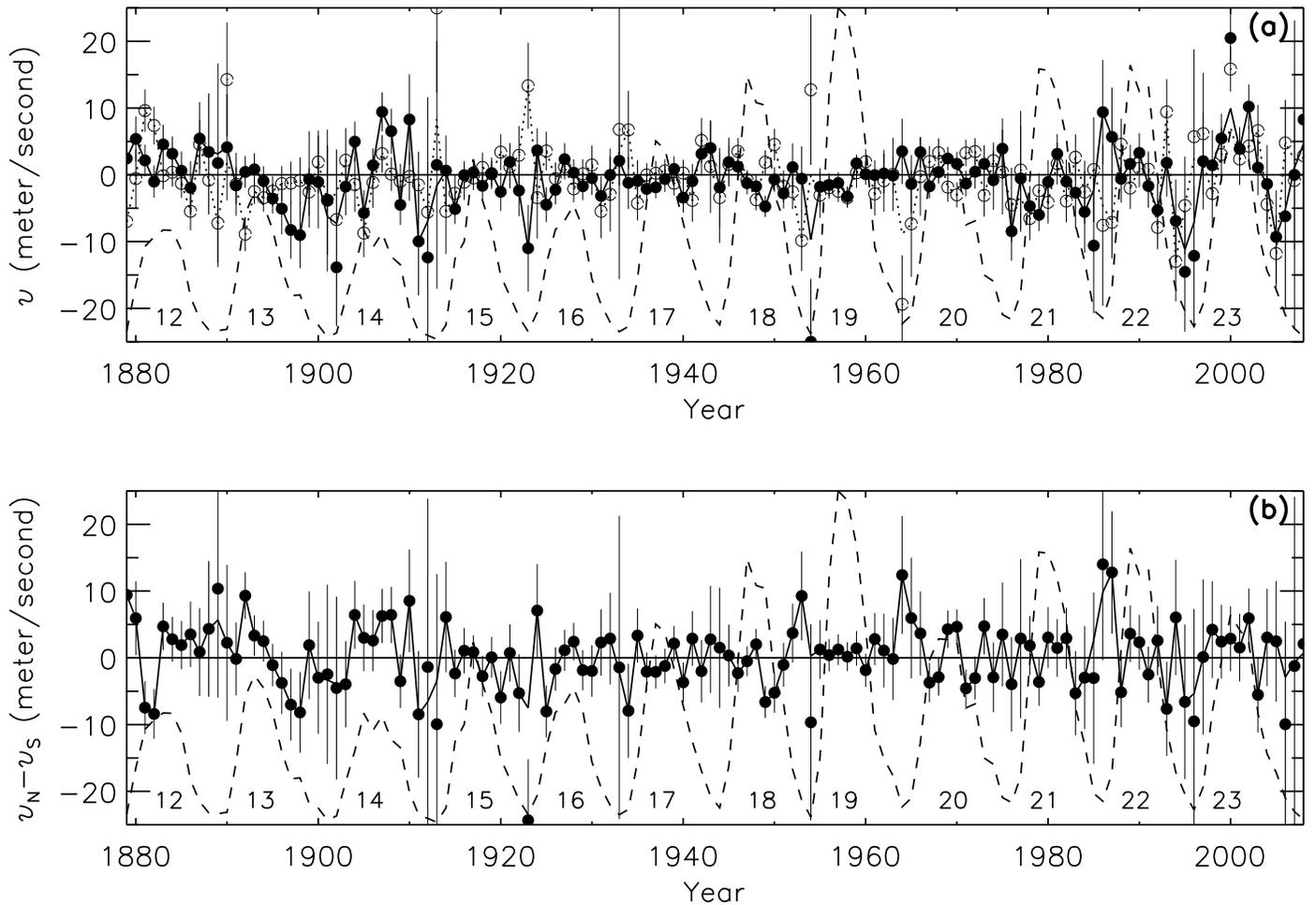}
\caption{Annual values of the mean meridional  motion 
of the sunspot groups during the period 1879\,--\,2008, (a) in the  northern 
hemisphere (filled circle-solid curve) and 
the southern hemisphere (open circle-dotted curve) and  
 (b) the corresponding north-south difference ($v_N - v_S$).   
The unconnected points represent  the  values that  
 are more than 15 m s$^{-1}$ or that  
have the large uncertainty, i.e., standard error (standard deviation, 
for the north-south difference) $> 2.6$.
In (a) a positive value represents the northbound motion. 
In both (a) and (b) the dashed curve represents the 
  annual variation
in sunspot 
activity during 1879\,--\,2008. 
The  Waldmeier cycle number is specified  near the maximum 
epoch of each cycle.} 
\end{figure*}

\begin{figure*}
\centering
\includegraphics[angle=90,width=\textwidth]{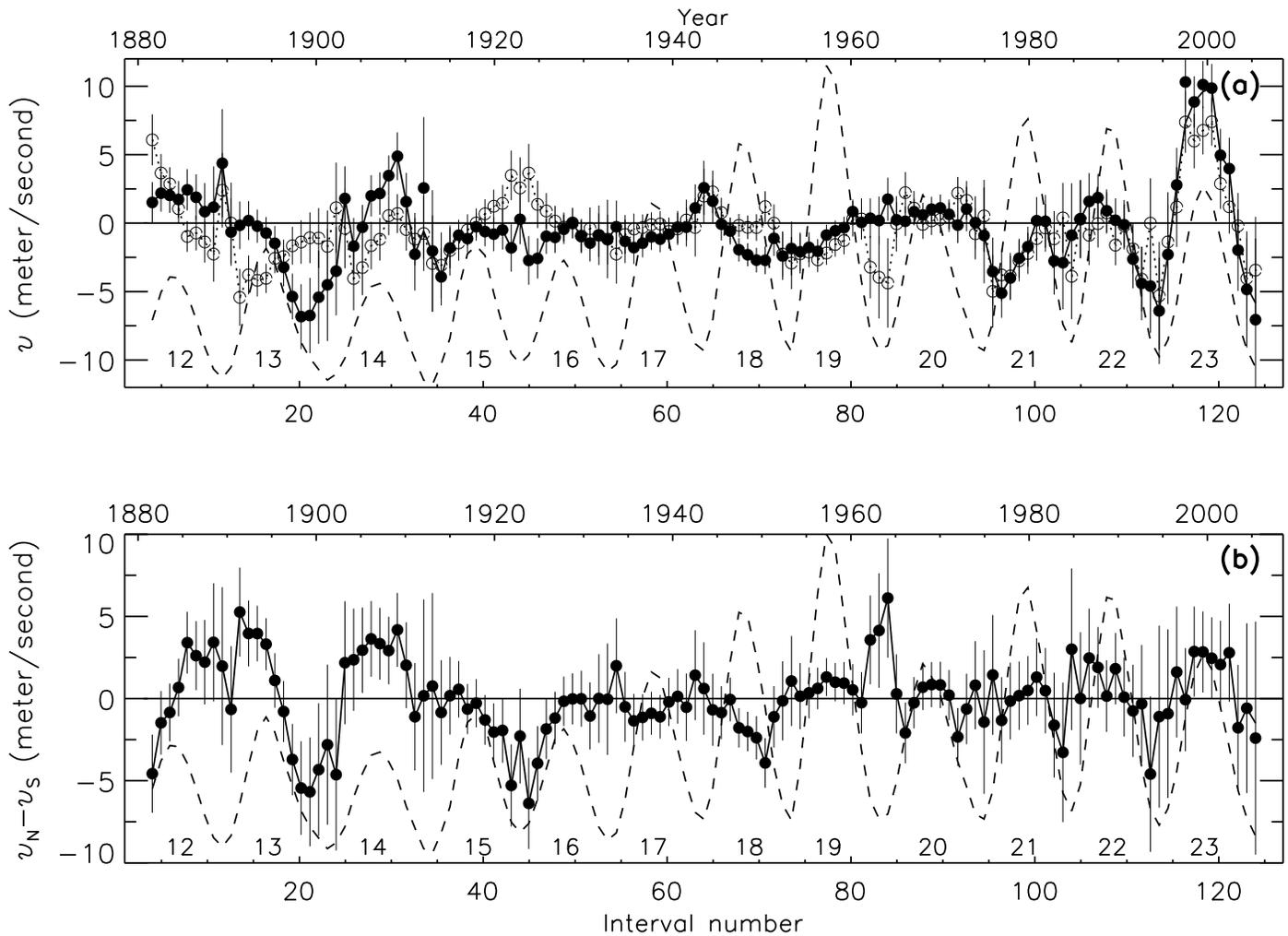}
\caption{Same as the Fig.~1, but  
determined from 4-year MTIs, 1879\,--\,1882, 1880\,--\,1883, ..., 
2005\,--\,2008. Here  
the unconnected points represent  the  values that 
 are more than 10 m s$^{-1}$ or  that  
have the large uncertainty, i.e., standard error 
(standard deviation, in the case of the north-south difference) 
 $> 2.6$, and 
the
 dashed curve represents the 
  variation in the sunspot number 
smoothed by taking 4-year running average.}
\end{figure*}

\begin{figure*}
\centering
\includegraphics[angle=90,width=\textwidth]{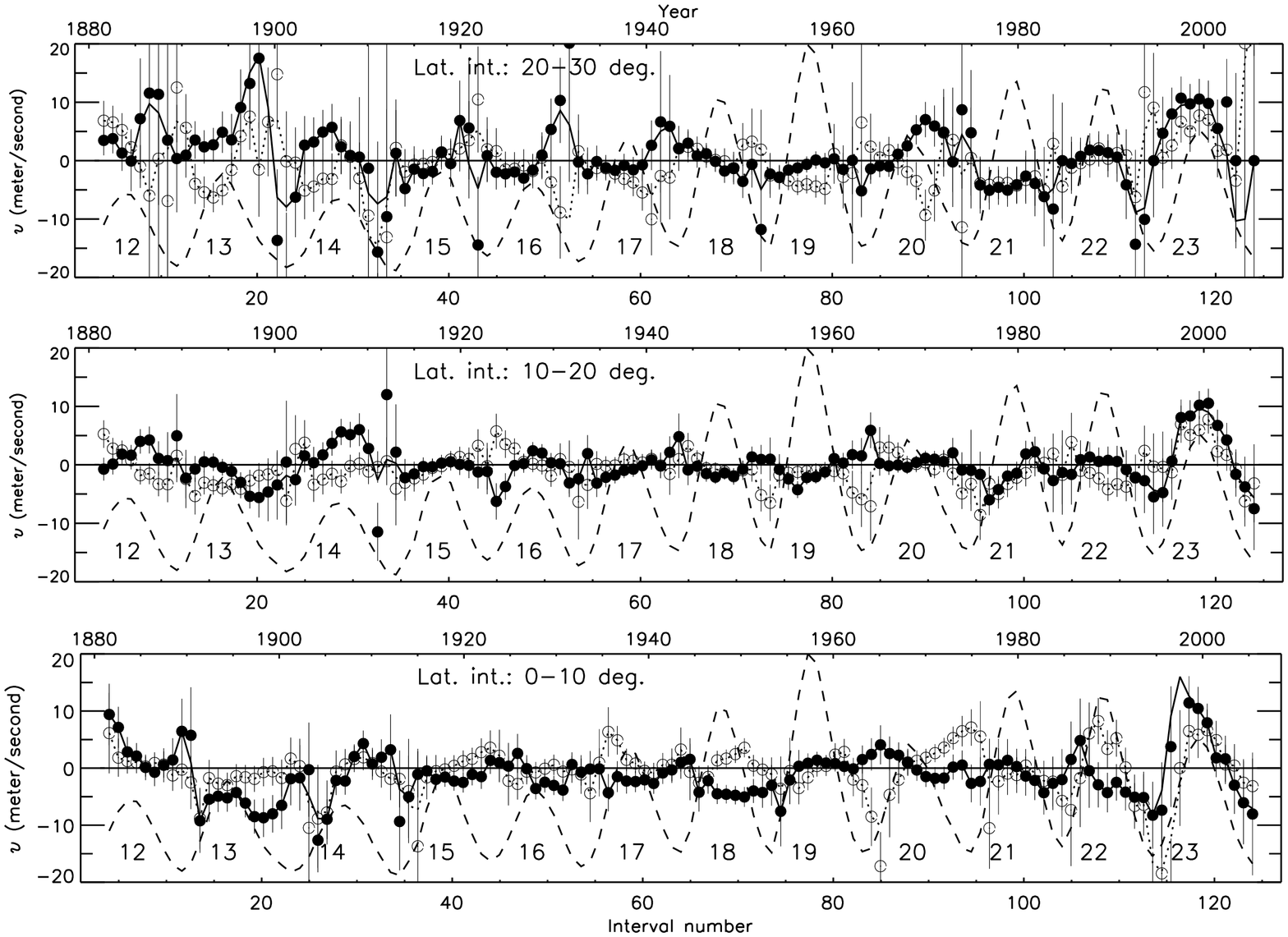}
\caption{Same as the Fig.~2(a), but  
determined 
 separately for different 10$^\circ$ latitude intervals.} 
\end{figure*}

\subsection{Temporal variations} 
Figure~1 shows the variations in the mean meridional  motion
of the sunspot groups in the northern and the southern hemispheres  and 
the corresponding north-south difference,  determined from the  
yearly spot group  data during 1879\,--\,2008. 
Figure~2 is the same as Fig.~1, but determined from  
the data in 4-year MTIs (3-years MTIs series is not shown because it is 
 found to be almost the same as that of 4-year MTIs series).  
Figure~3 shows the variations in the mean meridional motion of the spot groups 
in different 10$^\circ$ latitude intervals determined from the 
data in 4-year MTIs. To study the solar cycle 
variations in the mean meridional motion, we  also show the variations in the 
 sunspot activity in all  these figures.

As can be seen in  
Fig.~1(a), in the case of  
the original annual series,  the error bars are large. 
(The error bars 
are at the  
1$\epsilon$ level, where $\epsilon = \sigma/\sqrt{k}$ is the standard 
error,  $\sigma$ and $k$ are the standard deviation and the number of data 
points in a given interval, respectively.) In addition, 
there are some large spikes in the time series   during
 the minima of some cycles. We 
 therefore    applied 
the corrections to those  
  values whose $\epsilon$ values exceeded  2.6 times 
(correspond to 99\% confidence level) 
the corresponding median values or to those values $>$ 15 m s$^{-1}$; 
$i.e.$, these values are  replaced   with 
the average  of the corresponding values and 
their respective  two neighbors. (For the  beginnings of the 
time-series
  it is the average of 
 the values in the intervals 1 and 2 and in the endings    
 it is the average of the  values in the intervals $n-1$ and $n$.)
 In  the north-south differences ($cf.$, 
 lower panels of  Figs.~1 and 2), 
the error bar represents the   1$\sigma$ level 
 ($\sigma = \sqrt{\epsilon^2_1 + \epsilon^2_2}$, where 
 $\epsilon_1$ and $\epsilon_2$ represent the standard errors of the
 mean values
of the  northern 
and southern hemispheres, respectively).   
As can be seen in Fig.~2, in the case of the variations determined from 
4-year MTIs (in case of Fig.~3 in the lower latitude intervals),  
 the size of a  error bar 
  is  small.  However, we also applied the aforementioned 
corrections to  the MTIs time-series. In these cases, the corrected
time series included 
the   values that are  $\le$ 10 m s$^{-1}$.
 We  used 15 m s$^{-1}$ limit in the case of the annual time series 
because
the amplitude of the variation  is much 
 larger than  in the  MTIs series, and  
 to exclude only  extreme  outliers in the time series.

In Figs.~1\,--\,3 we can see the following: 
\begin{itemize}
\item The  pattern of the solar cycle 
variation in the mean meridional motion of the spot groups is
  different during different solar cycles.
 At the maximum epoch 
(year 2000, see Fig.~1)
 of the current  cycle~23, the mean motion
is stronger than  in the last  100 years and  
northbound in all the latitude intervals 
in both the northern and the southern  hemispheres.
 (The mean meridional motion is strong and  
 poleward in the northern hemisphere
and strong and equatorward in the
southern hemisphere.)
In this cycle the overall pattern of the variation in the mean motion   
closely resembles  the shape of this cycle; $i.e.$, 
 there is
a  high correlation between the activity 
and the mean meridional motion.
\item Overall there is a suggestion that  the mean meridional motion of the 
spot groups 
varies considerably on  time scales of  5\,--\,20 years.  
  The maximum amplitudes of the mean variations in  the northern and the 
southern  hemispheres 
 determined from the annual data  is about 10\,--\,15 m s$^{-1}$. 
The  amplitude of the mean variations 
determined from the 4-MTIs is on average about 5  m s$^{-1}$ and 
  seems to be  
 slightly large in the  individual 10$^\circ$ latitude intervals (see Fig. 3).
\item The difference 
between the mean motions in the northern and the southern hemispheres also 
varies on 5\,--\,20 year  time scales with  
the  maximum amplitude 5\,--\,10 m s$^{-1}$. However,  
the north-south difference is 
statistically
  significant  mainly  during 
the early cycles, 12\,--\,16, which are relatively weak ones, 
and mainly contributed from the differences in the motion of the high 
latitudes. The difference is negligibly small during the recent cycles. 
That is,  during the recent cycles, particularly in the 
current cycle~23,  the motion is highly symmetric about 
the equator.
\end{itemize}

Besides the above main features, in Figs.~1\,--\,3 
one can also see 
the following features: 
\begin{itemize}
\item Around the maxima of the large amplitudes cycles (e.g.,
 17, 18, 19, and  21), the average  motion
 in the whole Sun seems to be
slightly southward, whereas for  weak  cycles
the motion is either not significantly different from zero
(cf., cycles 16 and 20), or very 
different from zero and has  
northward direction (cf., 12, 14 and 23).
 During the  minima  of several  
 cycles, the motion
seems to be  strongly southward, and
 in some cycles   (cf., 16 and 20) it seems to go in opposite directions 
in the northern and the southern
hemispheres.
\item The behavior (changes in the  phase seem to be  
 taken place)
 of the meridional
 motions
  is  very different between the northern and the southern hemispheres  
 during 
the start of  cycles~16 and 20 (see Fig. 2), which are weak cycles that  
are four  cycles apart. 
A 44-year cyclic behavior is seen in the sunspot  activity
 (Javaraiah 2008).
\item  Near the maxima  of the much longer cycles (e.g., 13 and 23)
  in the high latitudes 
of both  the northern and southern hemispheres, the mean motion is largely
  northbound. 
 Javaraiah \& Ulrich (2006) found a positive correlation between the 
mean meridional motion of the spot groups of a cycle, 
mainly in the $20^\circ - 30^\circ$  
latitude interval  of the northern hemisphere, 
 and the length of 
the same cycle. In Fig.~3 the pattern of 
 the variation in 
the mean meridional motion of the spot groups in the $20^\circ - 30^\circ$ 
latitude interval of the northern hemisphere  
is largely consistent with this  result. 
\end{itemize}

\subsection{FFT power spectra}
Figure~4 shows the FFT power spectra of the corrected  annual time series 
of the mean meridional motions of 
the spot groups in the northern and the southern  hemispheres
 and that of the corresponding north-south difference, during 1879\,--\,2008.
 Figure~5 shows the
FFT power  spectra (low-frequency sides) of the mean meridional 
motions of the spot groups in 4-year MTIs.
It should be noted here that, in the FFT power spectrum of the data 
binned in  the  longer intervals (e.g. 4-year MTIs), 
 the peaks corresponding to the
high-frequency side are washed out and the  low-frequency peaks
are became broader.

\begin{figure}
\centering
\includegraphics[width=8.5cm]{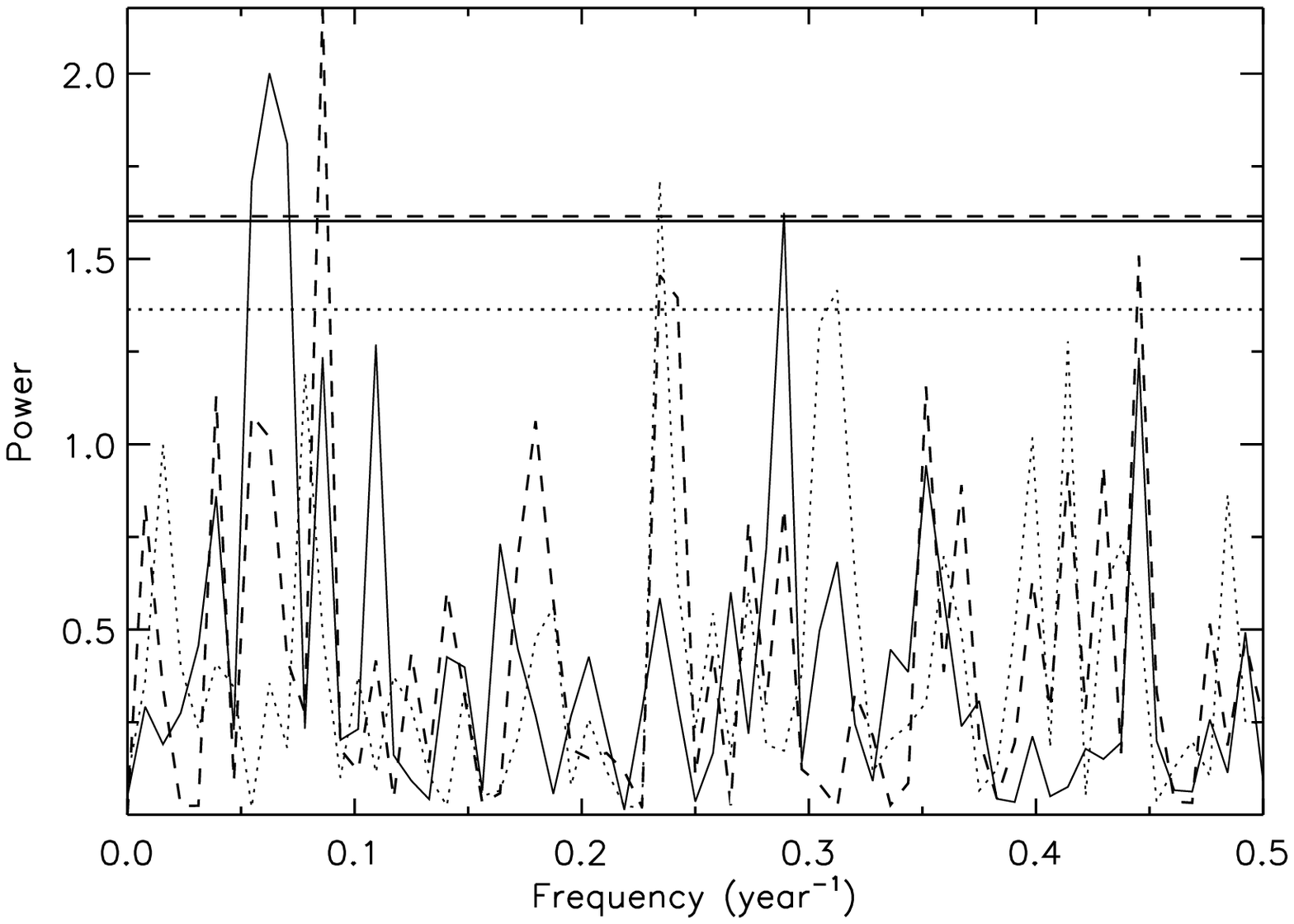}
\caption{FFT power spectra 
  of   the corrected yearly data 
 of the mean  
meridional motion of the sunspot groups in the northern
 hemisphere (solid curve)
 and the southern hemisphere (dotted curve), and 
 of the corresponding north-south differences (dashed curve).  
 The solid, dotted and  dashed
horizontal lines represent the 99\% confidence levels ($> 2.6 \sigma$ levels)
 of the
power in the  corresponding spectra represented by  the  solid, dotted and
 dashed curves, respectively.}
\end{figure}

As can be seen in Fig.~4, there are several  peaks in  
each of  the FFT spectra of the yearly data 
of  the 
mean meridional motion of the spot groups. 
 However, only the following peaks 
are significant on more than 99\% confidence level ($> 2.6\sigma$ levels):
In   the  spectrum of 
 the northern 
hemisphere, 
 the  peak at frequency 
$f  \approx 0.0625$ year$^{-1}$ (period $T \approx 16$ years); 
in the spectrum 
  of the southern hemisphere, 
the two peaks of relatively high frequencies which are closer to
 $f = 0.232$ year$^{-1}$ ($T = 4.3$ years) and 
$f = 0.312$ year$^{-1}$ ($T= 3.2$ years); 
and 
in the  spectrum of the north-south difference,
the  peak at frequency 
$f  \approx 0.0625$ year$^{-1}$ (period $T \approx 16$ years).
 In the spectrum of the northern hemisphere, the  peak at 
 $f \approx 0.286$ year$^{-1}$ ($T \approx 3.5$ years) is significant 
on a  99\% confidence level.  
As can be seen in Fig.~5, which shows the low-frequency sides of the 
 power spectra of the data  in the 4-year MTIs, in the spectrum of 
the     northern 
hemisphere the broad peak of the $\approx 16$ year period  is significant on 
 a 99\% confidence level. In the corresponding  spectrum of the southern 
hemisphere,  the peak at $f \approx 0.0158$ year$^{-1}$ 
($T \approx 63.5$ years)
is significant 
on a 99\% confidence level,
 and the peak at 
  $f \approx 0.079$ year$^{-1}$ ($T \approx 12.7$ years)
 is significant on a 95\%
confidence level. 
(A similar peak is also present
 in the yearly data.)  
 In the spectrum of  
 the north-south difference, a peak at $f = 0.394$ year$^{-1}$ 
($T = 25.4$ years) is significant on more than a 99\% confidence level, and 
the peak at $f \approx$ 1/18 year~$^{-1}$  
 significant on a 95\%
confidence level. 
Overall, all these results suggest 
that there are  around 5\,--\,20 years periodicities in 
the  mean meridional motion of the spot groups and that   there are 
  considerable differences in their levels of significance
 between
 the northern and the southern hemispheres. 
 
\begin{figure}
\centering
\includegraphics[width=8.5cm]{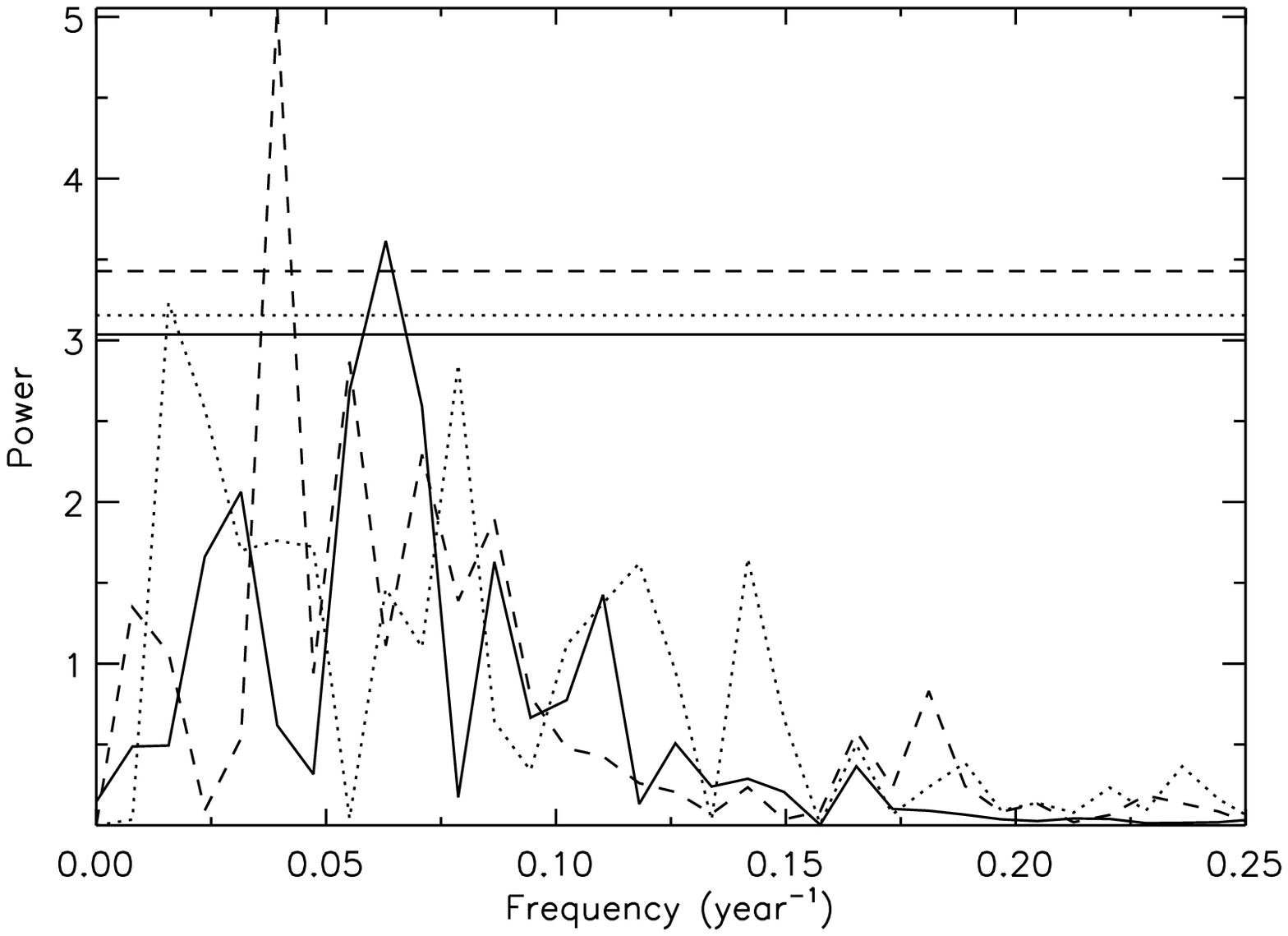}
\caption{The same as Fig.~4, but 
 for    the variations in the mean meridional motions of the spot 
groups in the 4-year MTIs (shown in Fig.~2).
 The values of the power only at the frequencies shown here are used
for determining the confidence levels  of the peaks
  in the respective spectra. The solid, dotted and dashed horizontal lines 
represent the 99\% confidence levels.} 
\end{figure}

 Figure~6 shows the FFT power  
spectra of the mean meridional  motion of the spot groups in different 
$10^\circ$ latitude
intervals of the northern and the southern  hemispheres during 4-year MTIs. 
As can be seen in these spectra,  there is  considerable latitude dependence
in the   periodicity of the mean motion of the spot groups in a hemisphere. 
For example, the  statistically significant $\approx 16$ year periodicity 
 found above in the mean motions of the spot groups in 
the northern hemisphere seems to contain  more contribution from the 
motions of the
 groups in $20^\circ$\,--\,30$^\circ$  latitude interval. Besides this, 
 the FFT spectrum of the mean motion in this latitude interval has a  
 strong peak around $f \approx 0.076$ year$^{-1}$ ($T \approx 13$ year).
In addition, there is  a suggestion that  this 
  (11\,--\,13 year) periodicity was strong in the mean meridional 
motion of the spot 
groups in the   
$20^\circ$\,--\,30$^\circ$ latitude interval of the northern hemisphere, 
whereas it was strong in the mean meridional motion of the spot groups in 
$0^\circ$\,--$10^\circ$ latitude interval of
 the southern hemisphere. 

 We would like to mention that 
most of the relatively high peaks in the FFT spectra
 seem to appear  at  
frequencies that  
  correspond to the integral multiples of one of the frequencies. 
Hence, they may look 
suspiciously like mathematical artifacts. However, they may not be 
mathematical artifacts since the length of the data used here is longer
 than double the longest
period that we find here. In addition,  
 before computing the FFT,  the mean value
  was subtracted  and a cosine
bell function was applied to both the first and the last 10\% of the time
series. These processes detrend the  time series and minimize the 
aliasing and leakage effects  (Brault \& White 1971). 
   The existence of 
the `harmonics' and 
`subharmonics'
may be a consequence of the Sun's behavior as a forced nonlinear 
oscillator (e.g., Bracewell 1988; Gokhale et al. 1992; Gokhale \& Javaraiah 1995). 

\begin{figure}
\centering
\includegraphics[width=8.5cm]{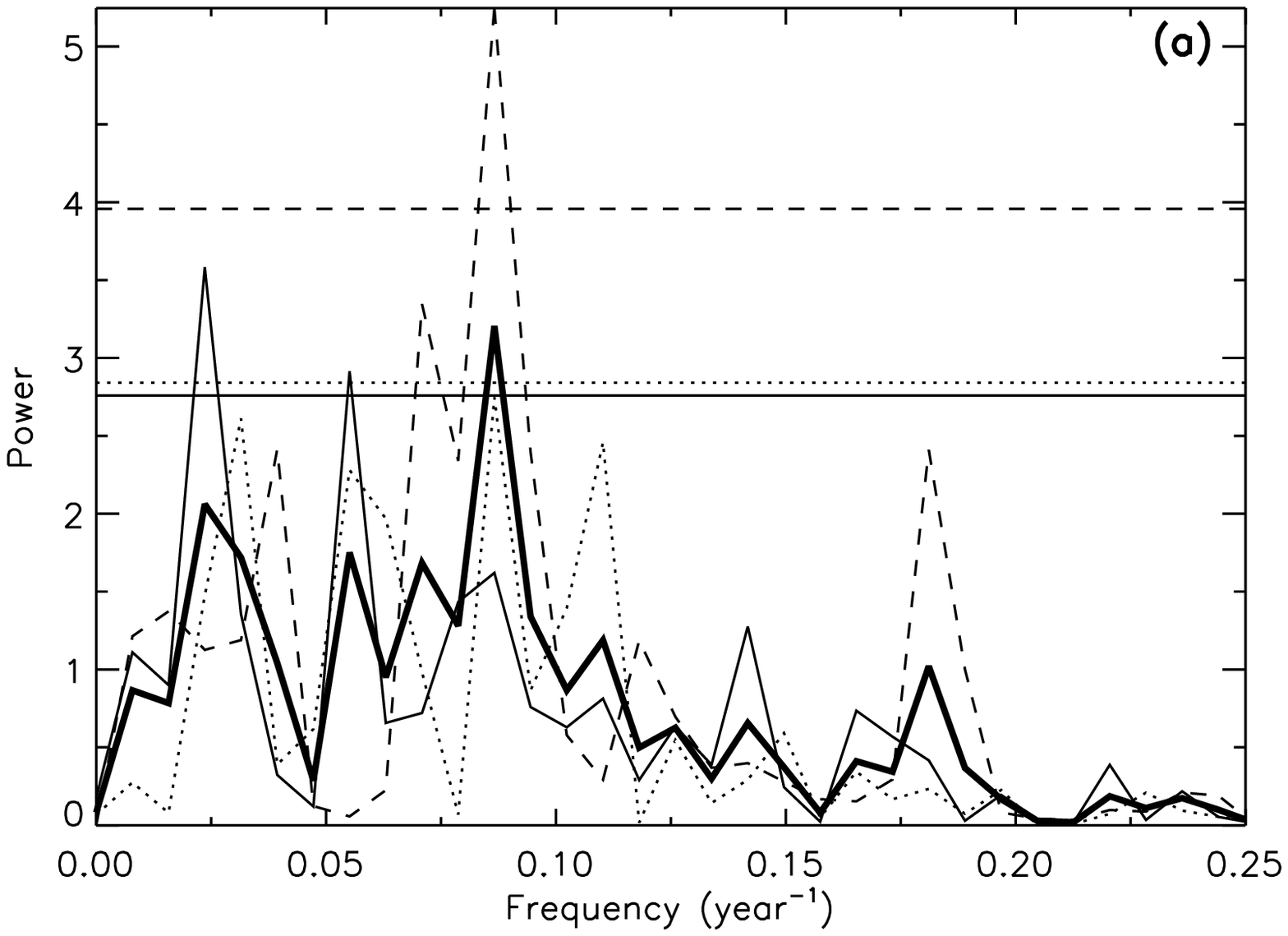}
\includegraphics[width=8.5cm]{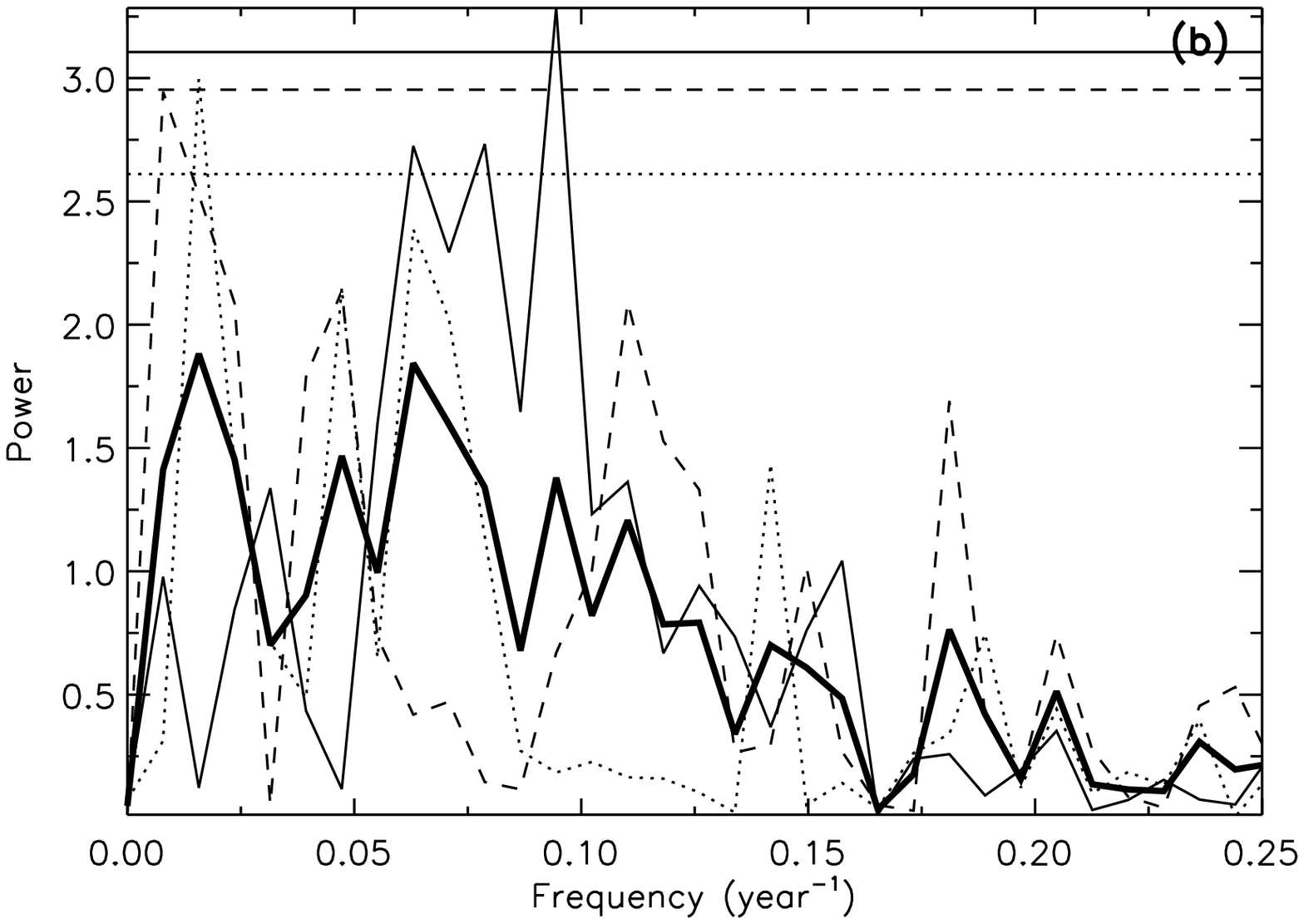}
\caption{FFT power spectra of the corrected variations 
      of the mean meridional motions of the spot 
groups in the different $10^\circ$ latitude intervals,
 $0^\circ$\,--\,$10^\circ$ (solid curve),  
 $10^\circ$\,--\,$20^\circ$ (dotted curve), and  $20^\circ$\,--\,$30^\circ$ 
(dashed curve),
 of   
 (a) the northern hemisphere (upper panel) and 
 (b) the southern hemisphere (lower panel),  
 during  4-year MTIs.  
  The solid, dotted, and dashed horizontal lines 
represent the 99\% confidence levels. 
 The values of the power only at the frequencies shown here are used
for determining the confidence levels  of the peaks
  in the respective spectra. The thick slid curve represents the mean spectrum 
of the spectra correspond to  the three latitude intervals.} 
\end{figure}

\subsection{MEM power spectra}
Figures~7 and 8 show the MEM power spectra of the 
 mean meridional motion of the spot groups determined from
the corrected data of the  annual and the 4-year MTIs  time-series.
As can be seen in these figures,  each  MEM spectrum  
shows a number of  well-defined peaks.  
The  MEM spectrum  
of  the annual data of the  southern 
hemisphere show 
the values 4.3-year and 3.2-year, and that of the north-south difference  
shows the value  11.9-year, for the corresponding significant
 periodicities found from 
the FFT analyses. 
The MEM spectrum of the 4-year MTIs  of the southern hemisphere shows 
the value 50-year for the 
 $\approx 63.5$ year peak found in the corresponding FFT spectrum.  
 The high significant broad peak of the   $\approx$ 16-year  
periodicity found  in the FFT spectra of the northern hemisphere data
 is  broken into 13.1-year and $\approx$ 18-year peaks 
in the corresponding MEM  spectra.
The 13.1-year  periodicity peaks are  well-defined in the 
 MEM spectra of  both the annual and the 4-MTIs time series
of the northern hemisphere data.
 In addition,   the MEM spectrum of the
4-year MTIs also shows a well-defined 
peak with 29.8-year periodicity.

\begin{figure}
\centering
\includegraphics[width=8.5cm]{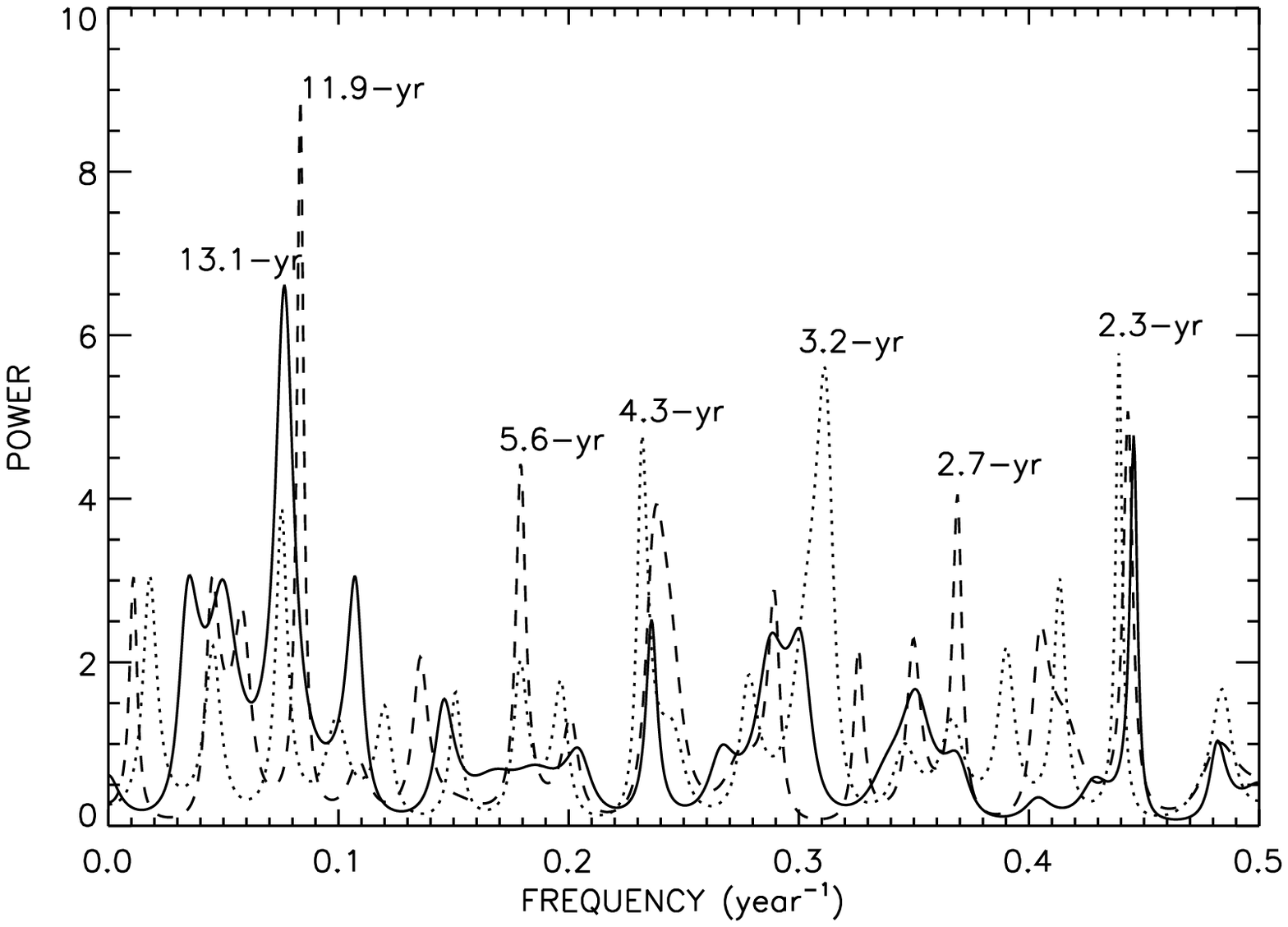}
\caption{MEM power spectra of the corrected yearly data of the mean 
meridional motions 
of the spot groups  in the northern hemisphere (solid curve) and
 the southern hemisphere (dotted curve), 
and  of the 
corresponding north-south differences (dashed curve). The 
values of the corresponding periods are marked near the considerably high  
peaks.}
\end{figure}

\begin{figure}
\centering
\includegraphics[width=8.5cm]{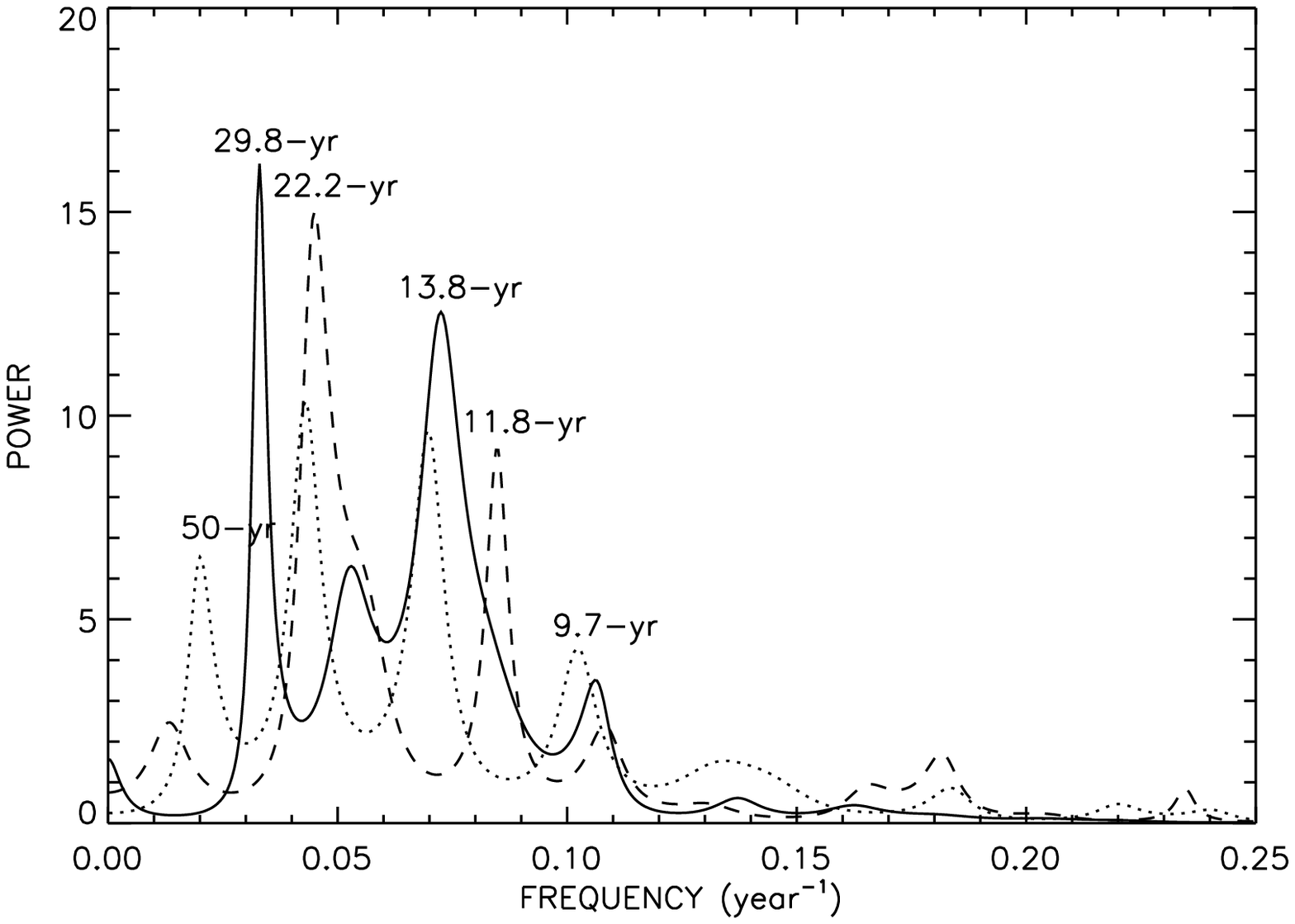}
\caption{The same as Fig.~7, but 
 for    the variations of the mean meridional motions of the spot 
groups in 4-year MTIs (shown in Fig.~2).}
\end{figure}

  The 3.5-year periodicity found 
in the FFT spectrum of the annual data of the northern hemisphere
 is also present in the corresponding 
MEM spectrum, but the corresponding peak 
 is  not clearly defined.
The peaks of $\approx$ 2.3-year and 4.3-year periodicities  are present 
in the MEM spectra  of all the three annual time series.   
The MEM spectrum of the annual data of the north-south 
 asymmetry also shows 
the peaks of 5.6-year and 2.7-year periodicities and the spectrum of 
corresponding 4-year MTIs shows  a  peak 
of 22.2-year periodicity.

\begin{figure}
\centering
\includegraphics[width=8.5cm]{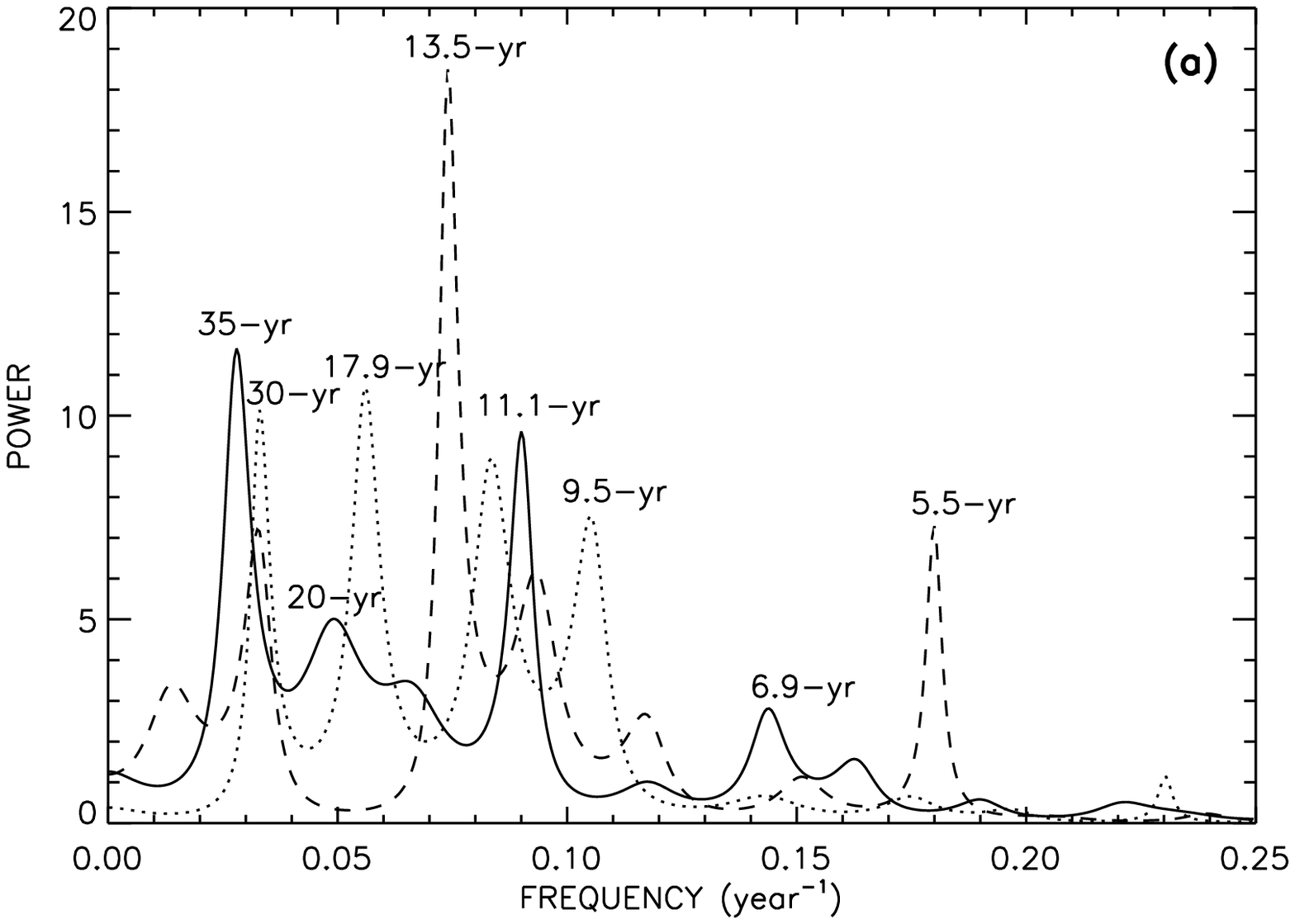}
\includegraphics[width=8.5cm]{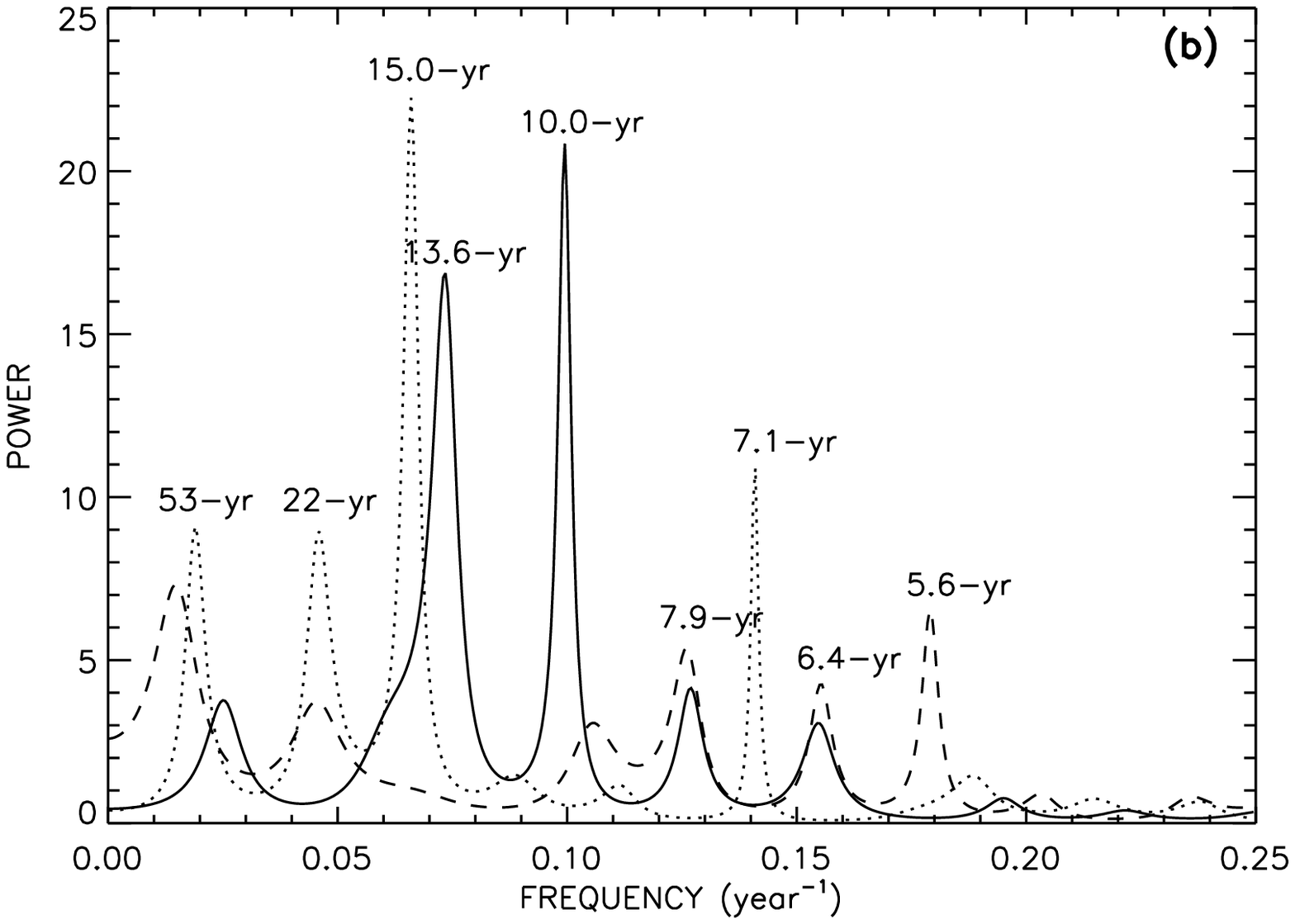}
\caption{MEM power spectra of the corrected variations 
      in the mean meridional motions of the spot 
groups in the different $10^\circ$ latitude intervals, 
 $0^\circ$\,--\,$10^\circ$ (solid curve),  
 $10^\circ$\,--\,$20^\circ$ (dotted curve), and  $20^\circ$\,--\,$30^\circ$ 
(dashed curve),
of (a) the  
northern hemisphere (upper panel) and 
 (b)  southern hemisphere (lower panel),  
 during  4-year MTIs.}
\end{figure}

Figure~9 shows the MEM spectra of the corrected 
data in the different $10^\circ$ 
latitude intervals of the northern and the southern hemispheres 
during the 4-year MTIs. As can be seen in this figure, 
the results from the MEM analysis  are  consistent with the results 
found from the FFT analyses and
 suggest that there is a strong
hemispheric and latitude dependency in the periodicities of the mean motion of
 the spot groups.

\subsection{Wavelet power spectra}
Figures~10 and 11  show the  Morlet wavelet power spectra, normalized by
 $1/\sigma^2$ (here $\sigma$ is the standard deviation of the 
concerned data sample), and the corresponding 
global spectra 
of the mean meridional motion of the 
spot groups determined from the
corrected annual and  4-MTIs time series, 
respectively. 
As can be seen in 
Fig.~10, during the period 1880\,--\,2007, the 3.2-year and 4.3-year 
periodicities  occurred  relatively   consistently in the 
mean meridional motion of the spot groups in the southern hemisphere.  
 This result is  highly  
consistent with the result that the aforementioned periodicities are found 
 statistically significant in the FFT analysis.
The 3.2-year periodicity and also 
 a  2.3-year periodicity were  prominent around 1990.
In the wavelet spectra, Figs.~10(b) and 11(b),
  of the mean motion of 
the spot groups in 
the southern hemisphere, there  
is a suggestion that the
dominant  periodicity  evolved slowly  
 with time, from $\approx$ 15~years to $\approx$ 30~years,  over the
period 1880\,--\,2007.  In Figs.~10(a) and 11(a)  there is a 
suggestion that   
the 13.1-year periodicity  and a $\approx$ 20-year periodicity 
 of the mean motion of the spot groups in
 the northern hemisphere are strong   
  after 1980 and before 1920. 
 The $\approx$ 30-year 
  periodicity  seems  to have occurred throughout 
the period 1880\,--\,2007, with more powerfully before 1920 and after 1960.

In  Figs.~10(c) and 11(c) there is an evidence that  the 
 11.9-year periodicity  occurred consistently in the north-south 
difference in the mean motion.  
This  is consistent with that the same  periodicity  
     is found to be  highly significant in the 
FFT analysis of the north-south difference.
The 29-year periodicity 
in  the north-south 
difference (found in the MEM analysis) was weak or absent after around 1970. 
In fact, the north-south 
difference is itself  very small and statistically
  insignificant during  
the recent cycles (see Figs.~1 and 2). 

\begin{figure*}
\centering
\subfigure{
\includegraphics[width=6.0cm]{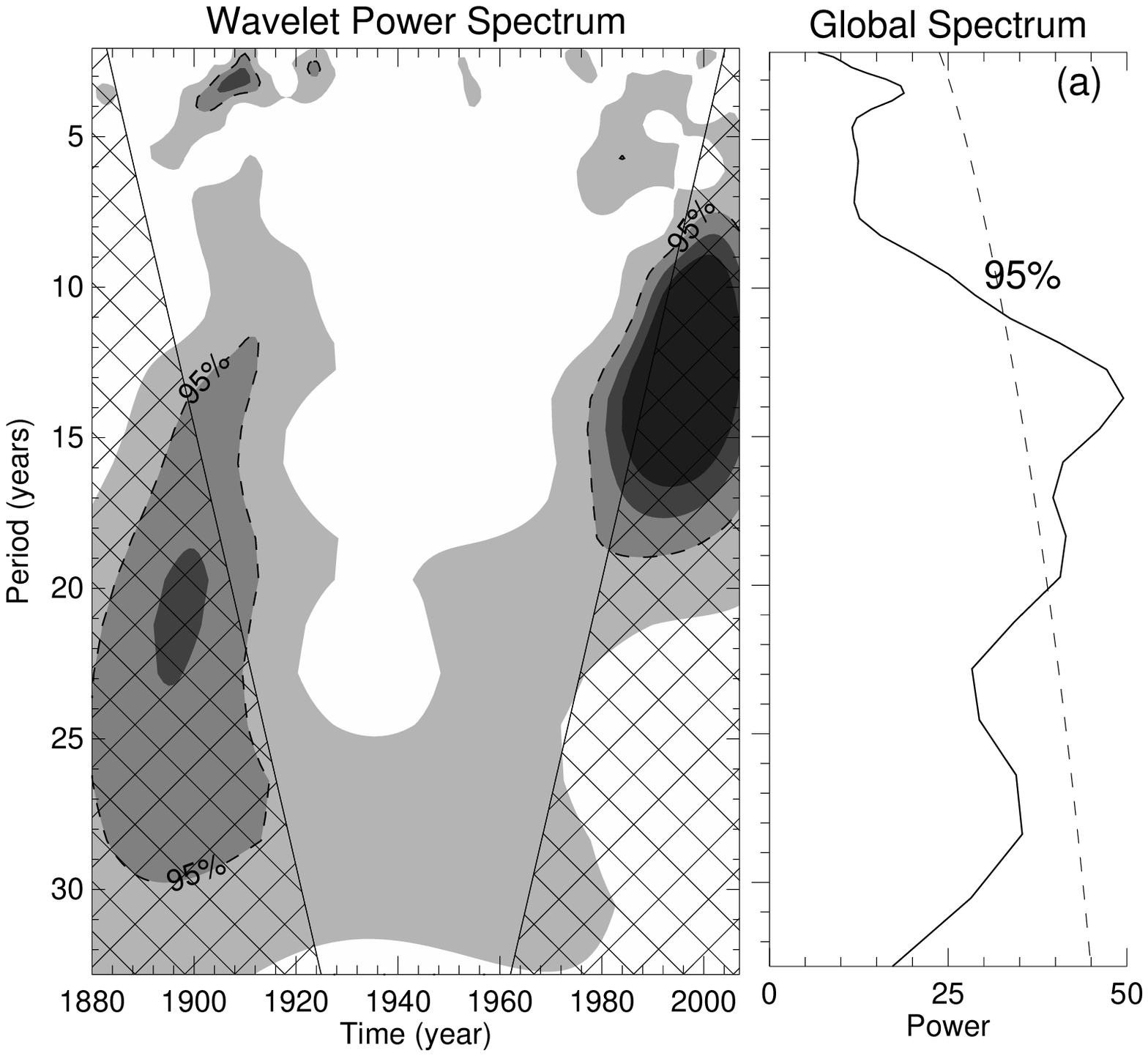}}
\subfigure{
\includegraphics[width=6.0cm]{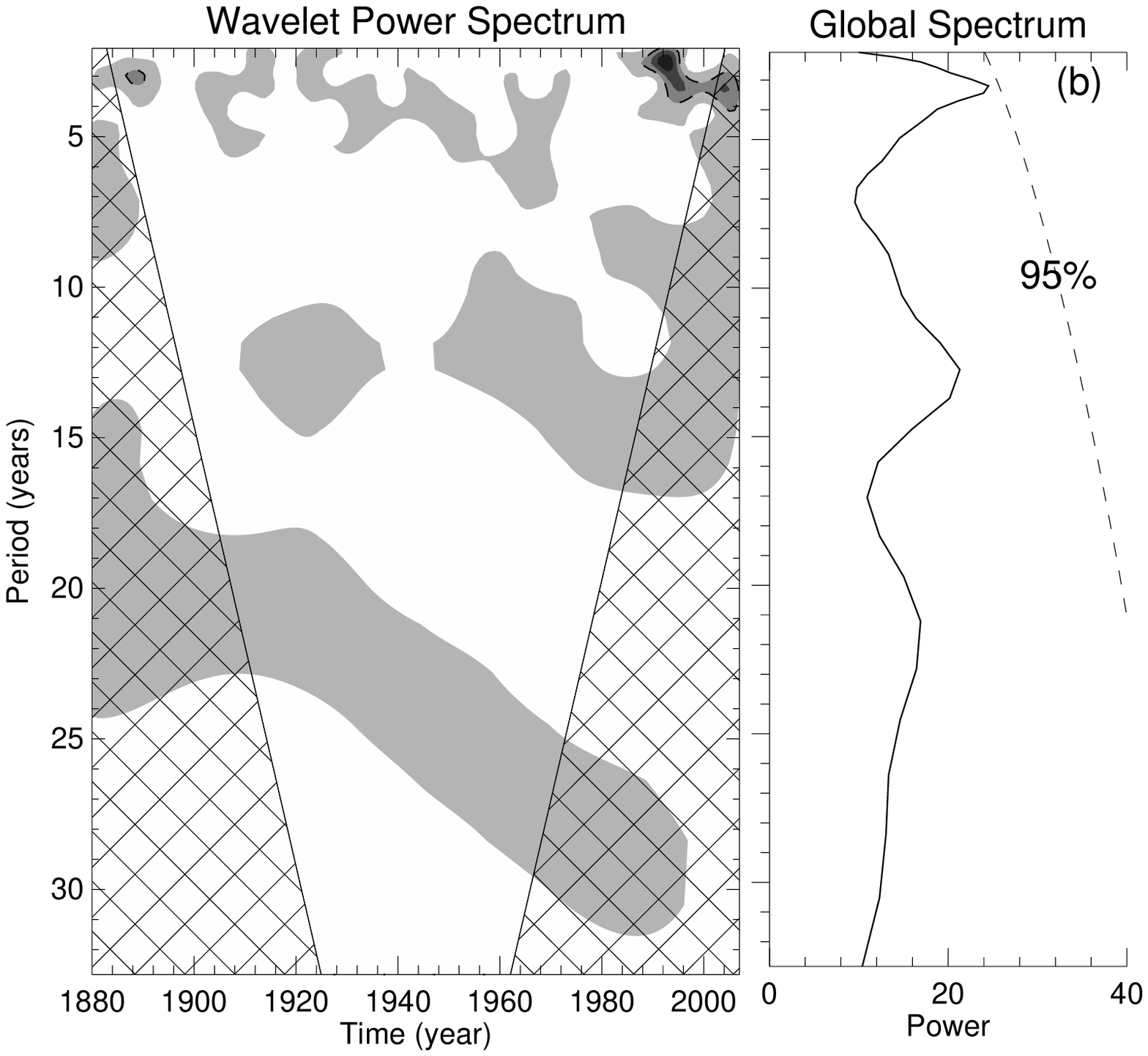}}
\subfigure{
\includegraphics[width=6.0cm]{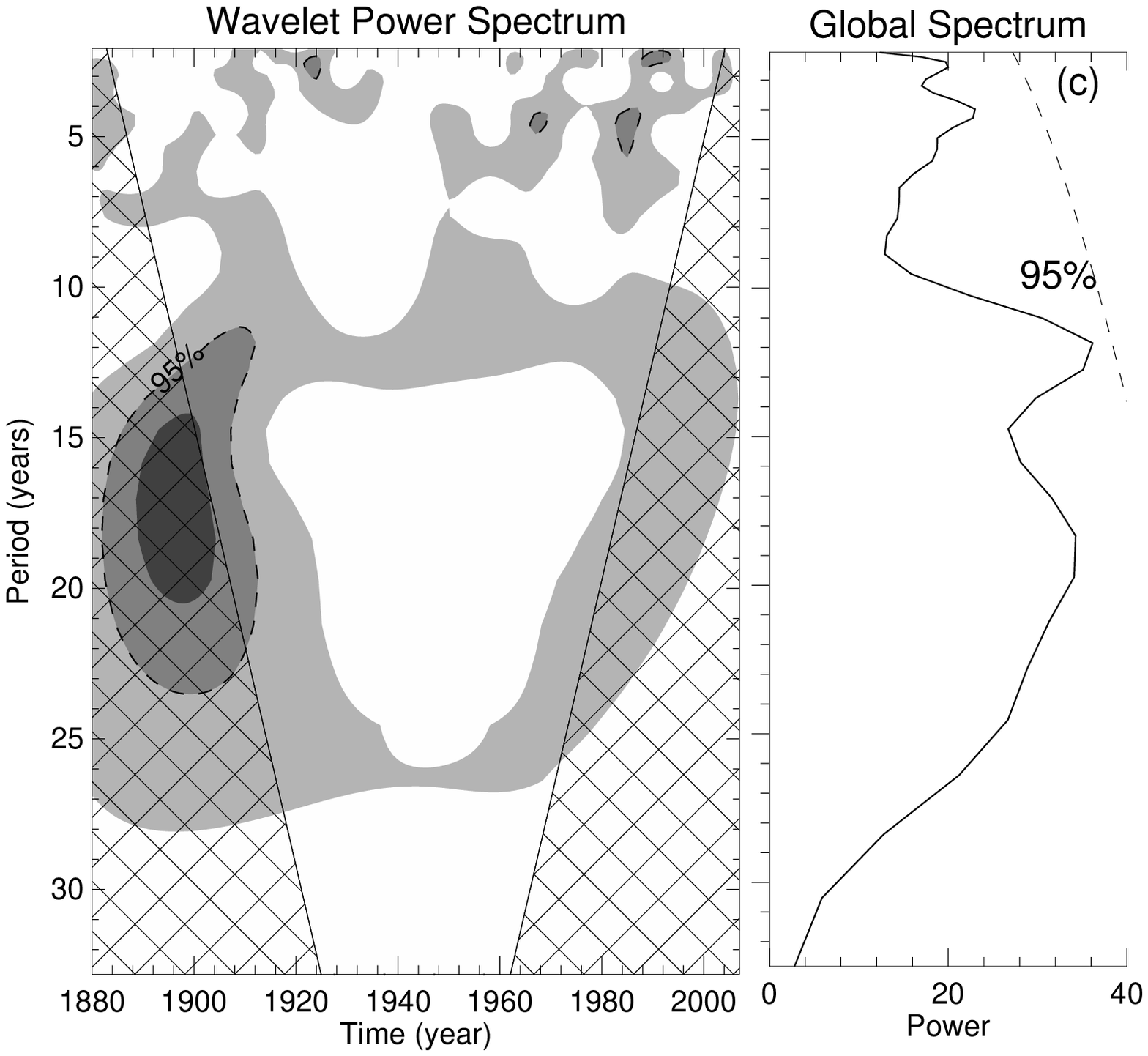}}
\caption{Wavelet power spectra and the global spectra 
of the corrected yearly data 
of the mean meridional 
motion of the sunspot groups  
  (a) in the northern hemisphere,  
  (b) in the southern hemisphere, and (c)  the corresponding  north-south 
difference.  The wavelet spectra are normalized by the variances of 
the  corresponding time series. 
The shadings are  at the normalized variances of 2.0, 3.0, 
 4.5, and 6.0.
 The dashed curves represent the 95\% confidence levels, 
 deduced by assuming a white noise process. 
The cross-hatched regions indicate the ``cone of
influence", where edge effects become significant (Torrence \& Compo 1998).}
\end{figure*}

\begin{figure*}
\centering
\subfigure{
\includegraphics[width=6.0cm]{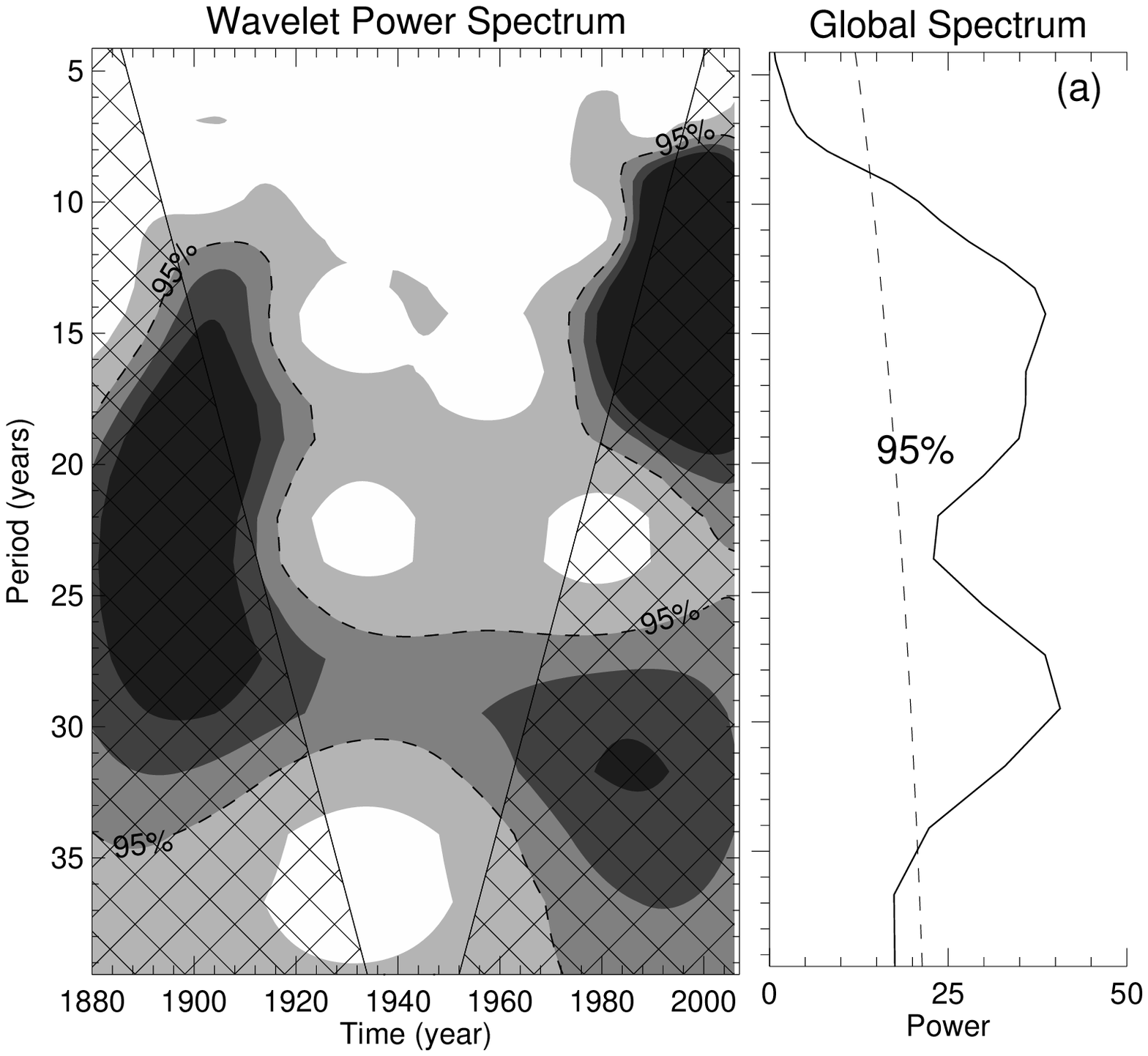}}
\subfigure{
\includegraphics[width=6.0cm]{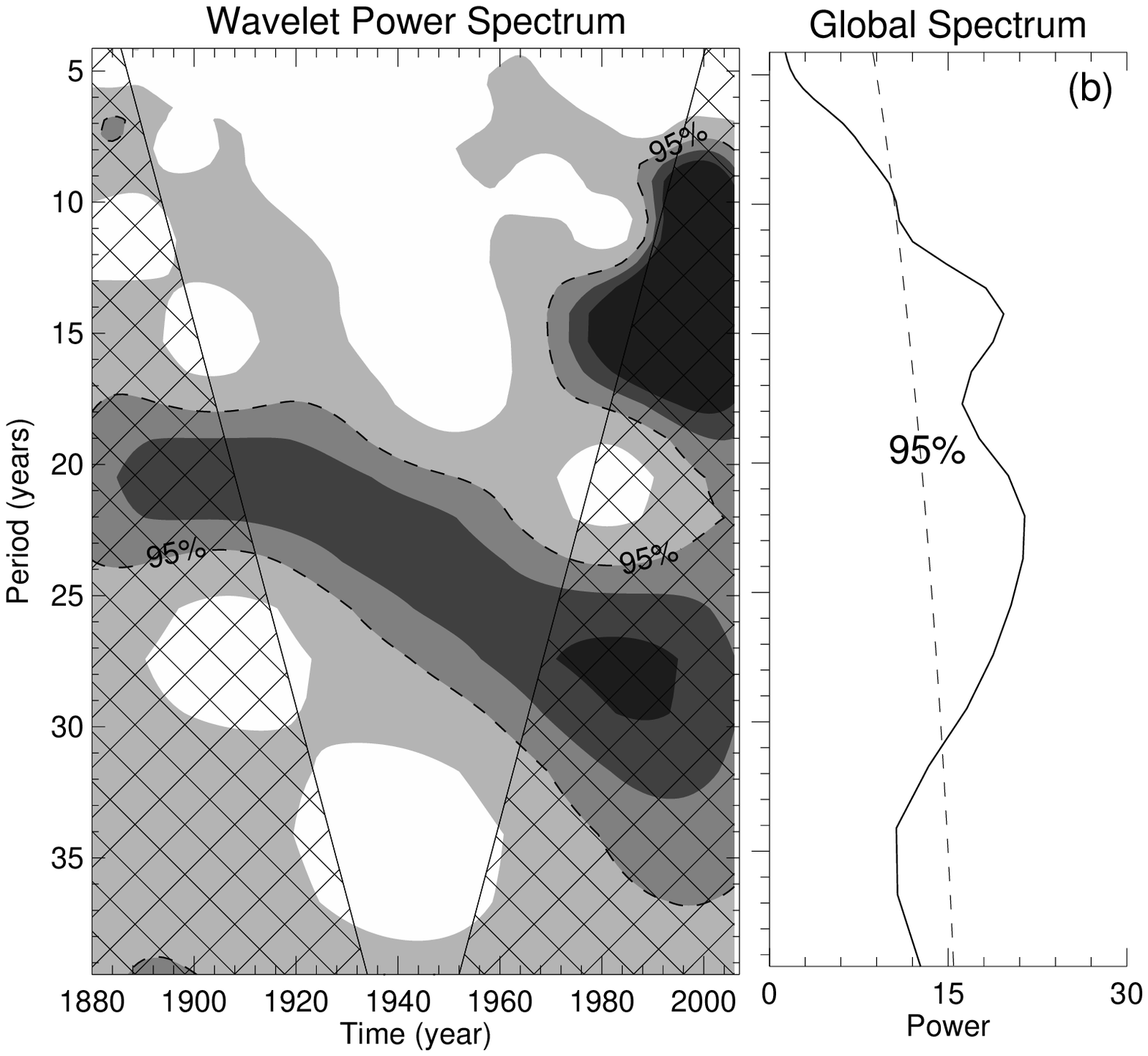}}
\subfigure{
\includegraphics[width=6.0cm]{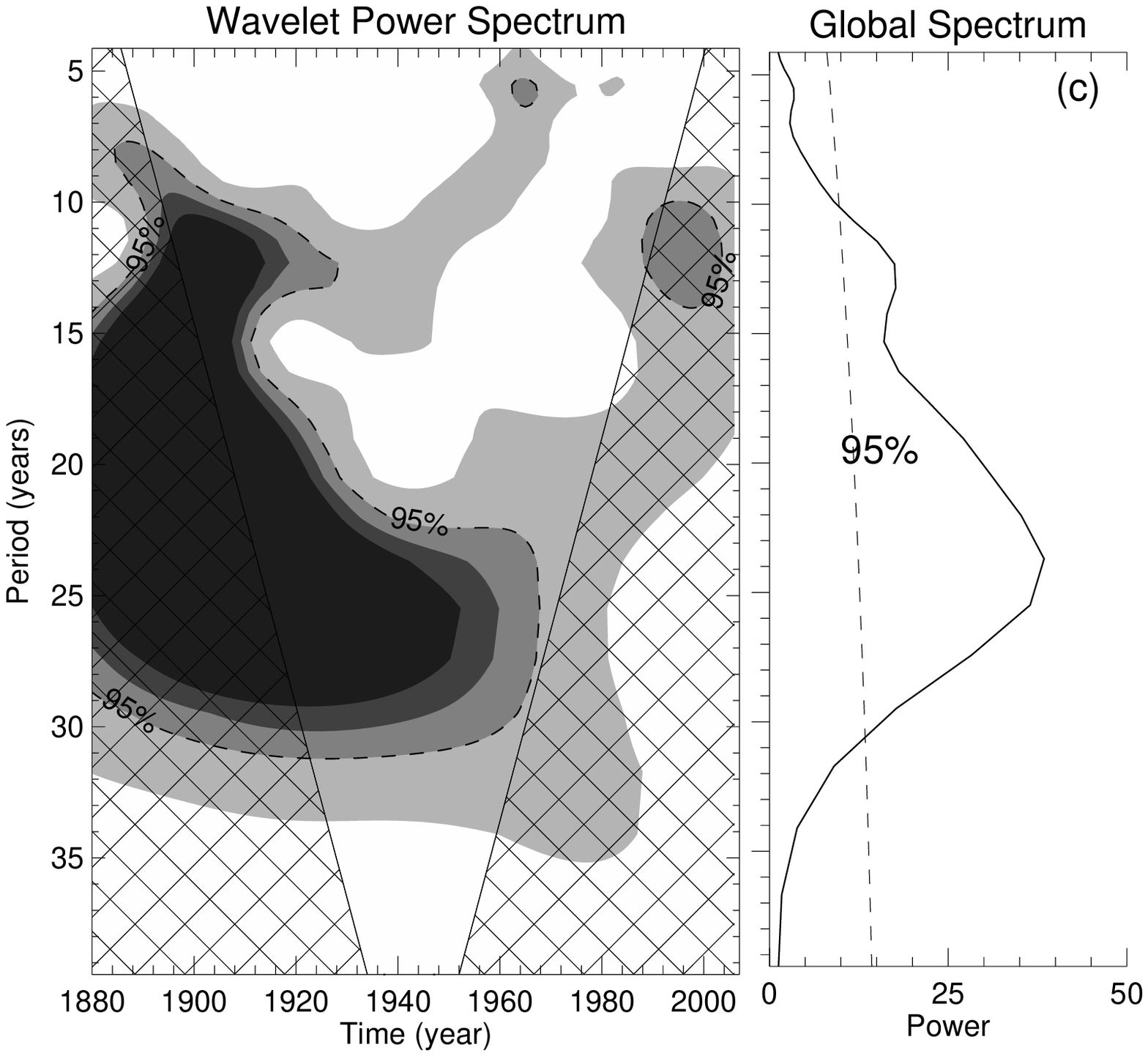}}
\caption{The same as Fig.~10, but for the variations 
in the mean meridional motion of the spot groups in 4-year MTIs.}
\end{figure*}

Figure~12 shows the wavelet power spectra of the corrected data in the 
different $10^\circ$ 
latitude intervals of the northern and the southern hemispheres, in  
 the 4-year MTIs.
 As can be seen in this figure, 
 there  is 
 a considerable latitude-time dependence in the 
  periodicities of the mean meridional motion of the spot groups. 
The $\approx$ 20-year and 
the $\approx$ 30-year periodicities
 probably exist  
in the mean meridional motion of the 
spot groups
of the lower latitudes of
 the northern hemisphere, throughout the  period 1880\,--\,2007. 
 The evolution of a $\approx$ 15-year periodicity to a $\approx$ 30-year 
periodicity, seen above in the  
spectrum of the mean motion of the whole southern hemisphere (see Fig.~10(b)), 
is not clearly visible in  the spectrum of the mean motion in 
any latitude interval of this hemisphere. On the other hand,  
there is  a suggestion that, in $10^\circ - 20^\circ$ latitude 
interval of the northern hemisphere,   a periodicity in the mean motion   
evolved from $\approx$ 16~years to $\approx$ 10~years 
over the
period 1880\,--\,2007. It was  the 
 opposite in the  $20^\circ$\,--\,$30^\circ$  latitude-interval.
In this latitude interval,  
the $\approx$ 40-year periodicity might have 
 evolved to a $\approx$ 25-year periodicity. 
Overall the wavelet analyses suggest 
that there is a considerable latitude-time dependency in the periodicities 
in the mean meridional motion of the spot groups.

\begin{figure*}
\centering
\subfigure{
\includegraphics[width=6.0cm]{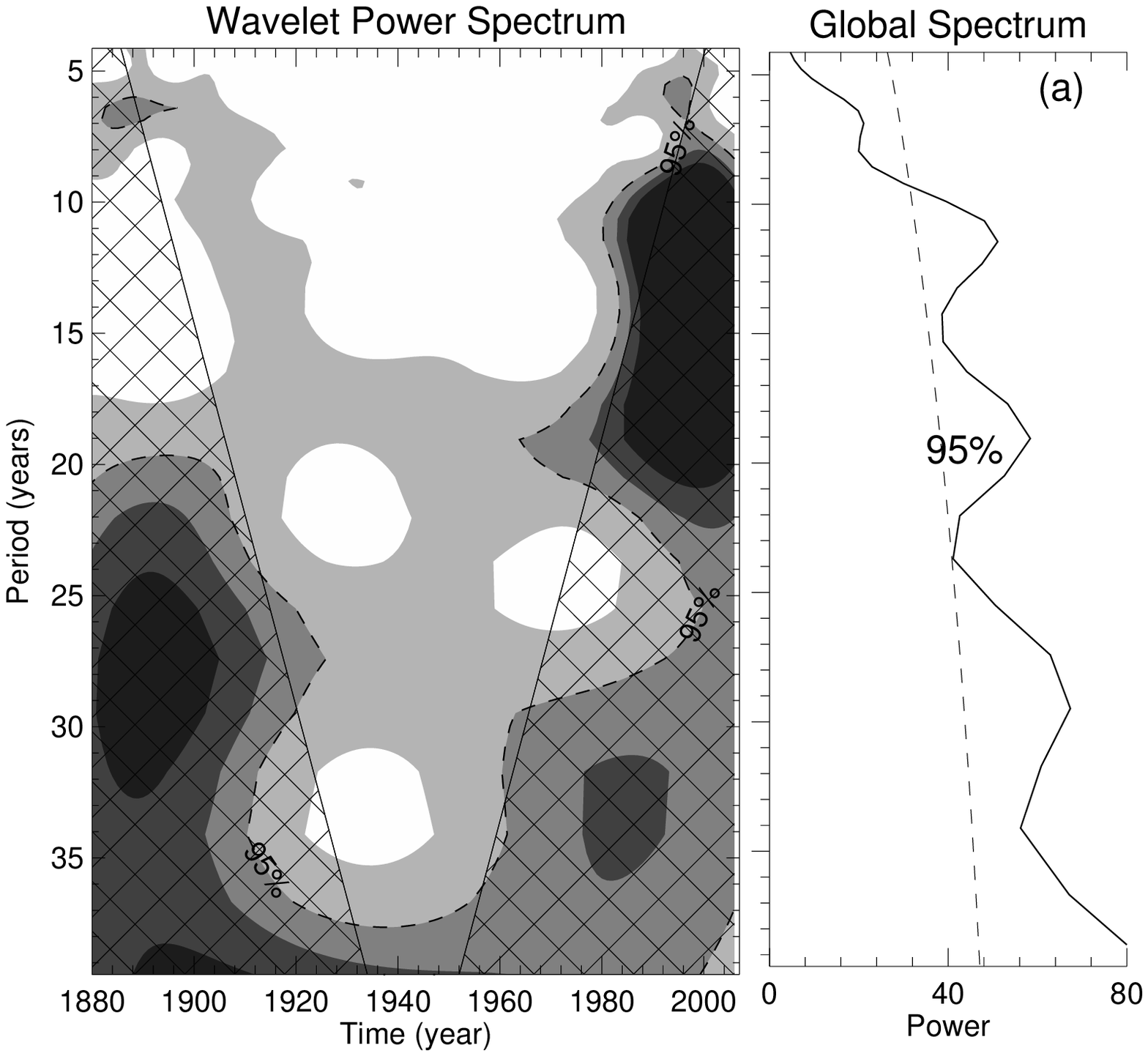}
\includegraphics[width=6.0cm]{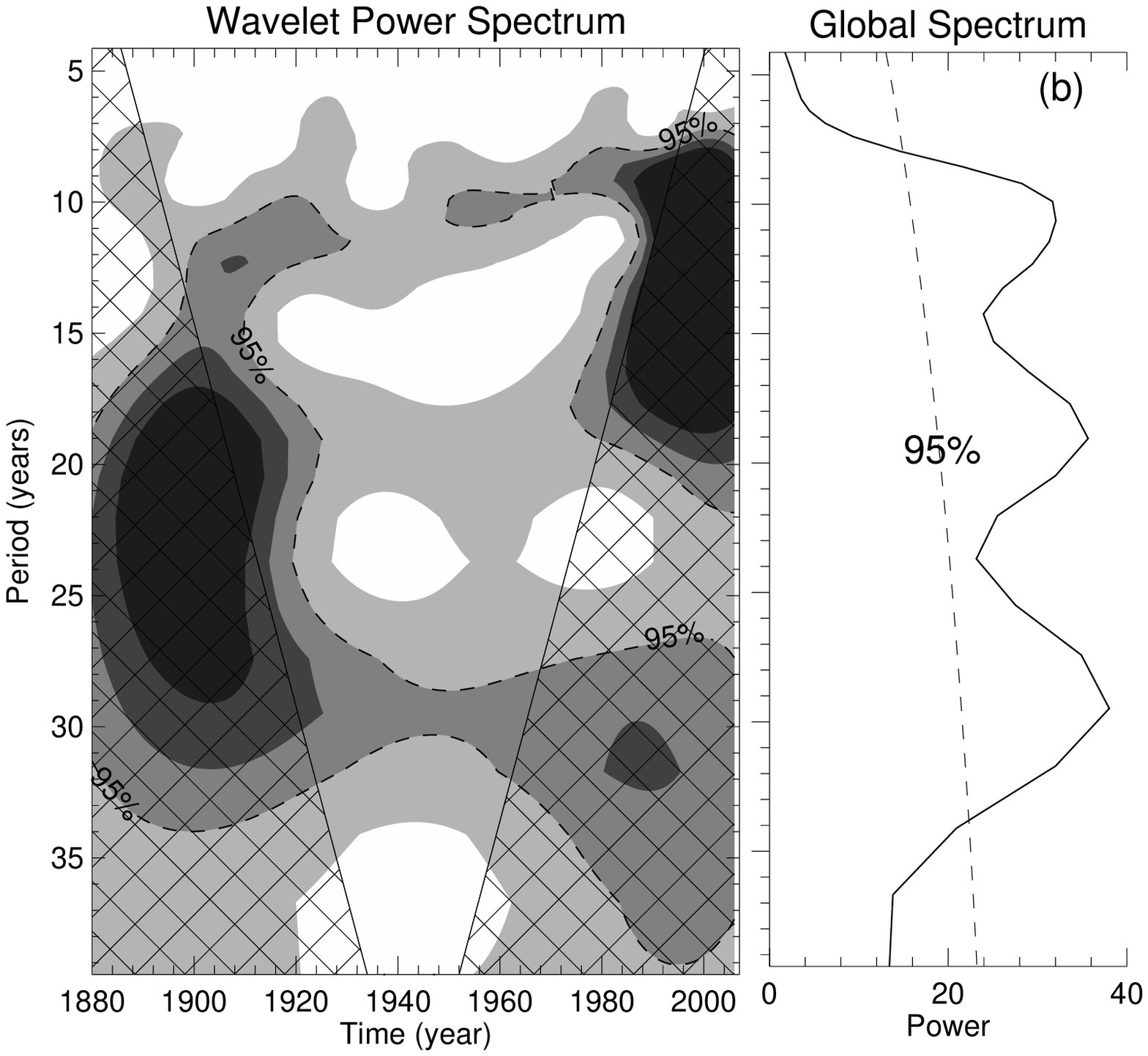}
\includegraphics[width=6.0cm]{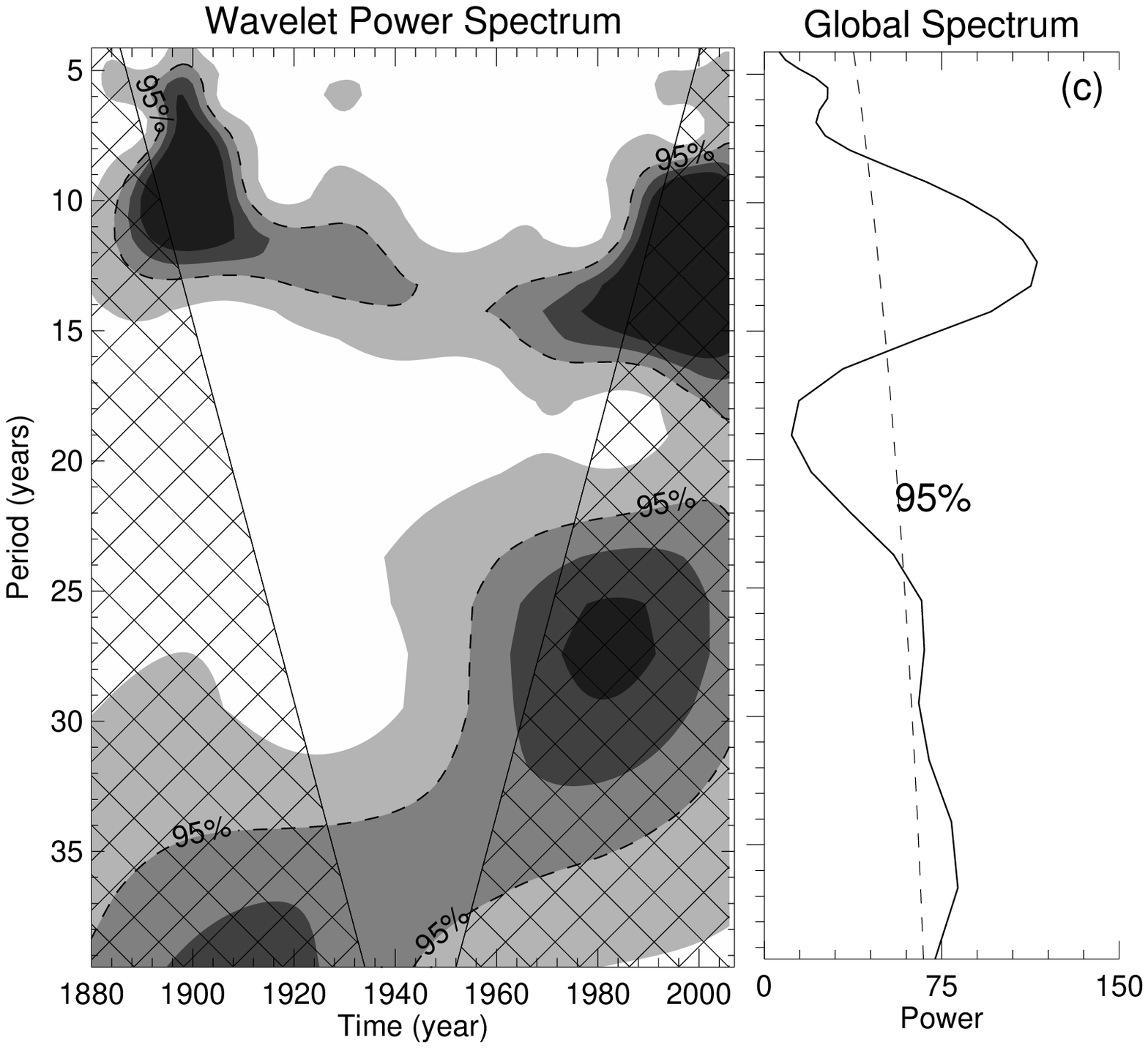}}
\subfigure{
\includegraphics[width=6.0cm]{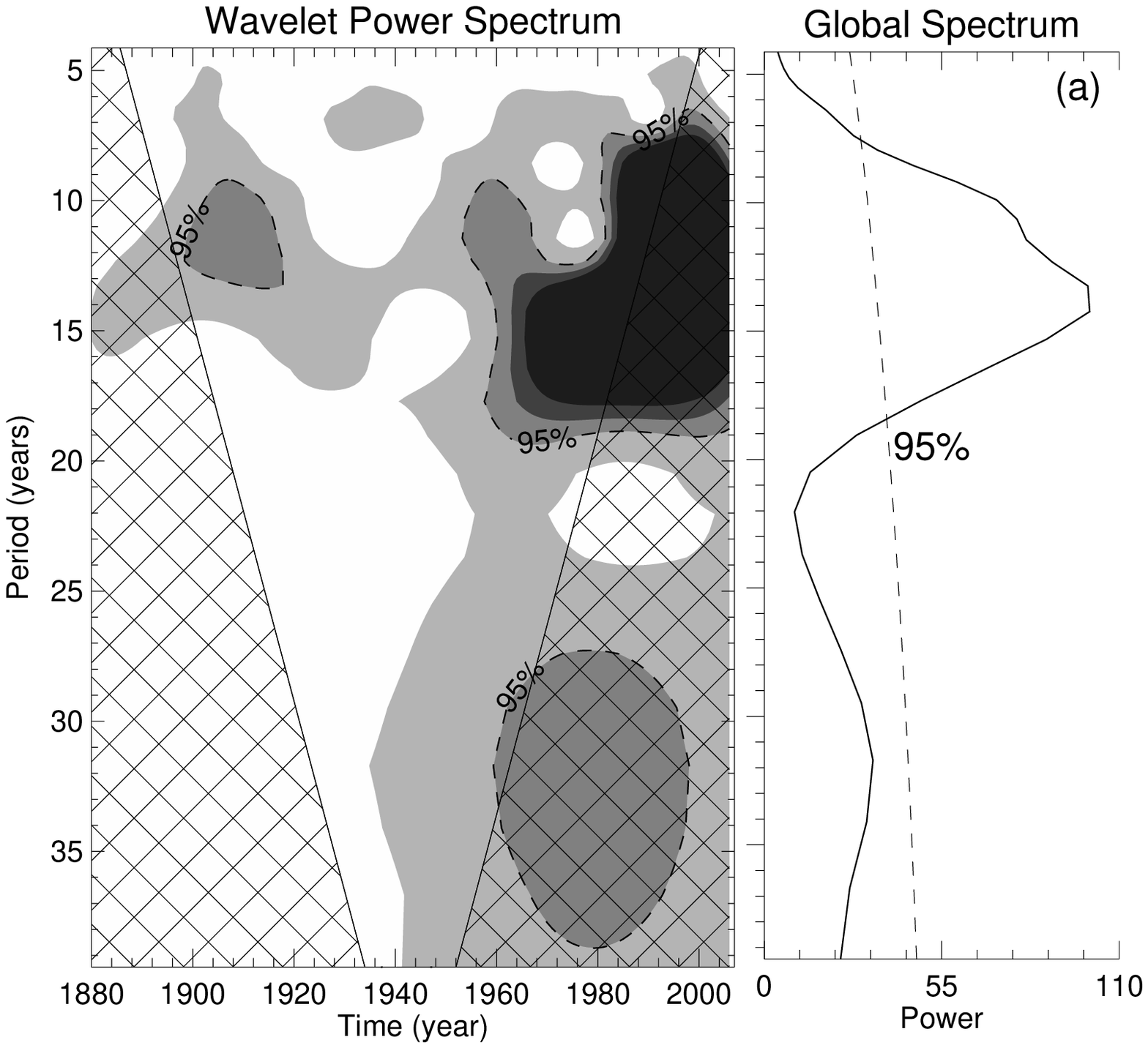}
\includegraphics[width=6.0cm]{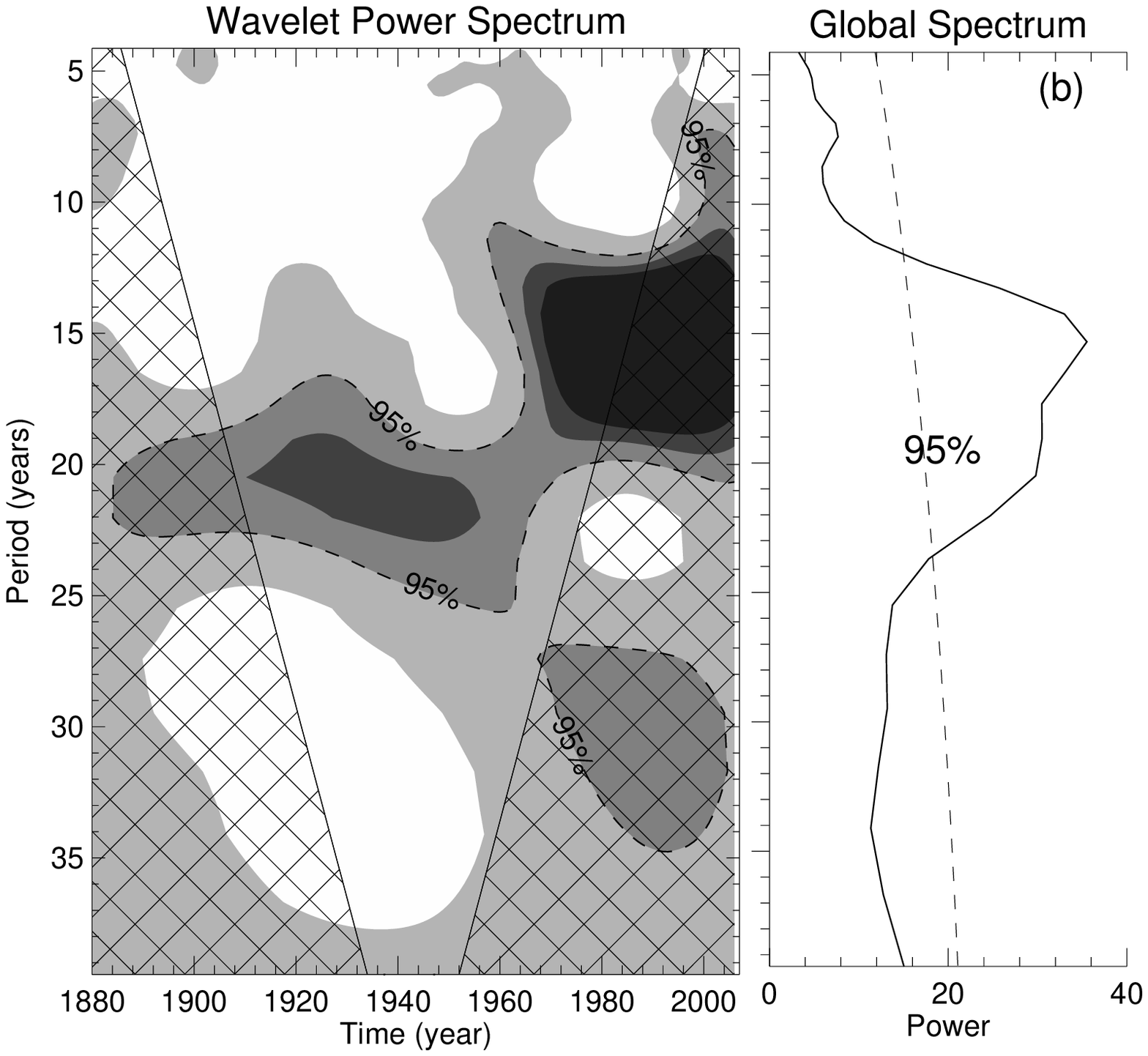}
\includegraphics[width=6.0cm]{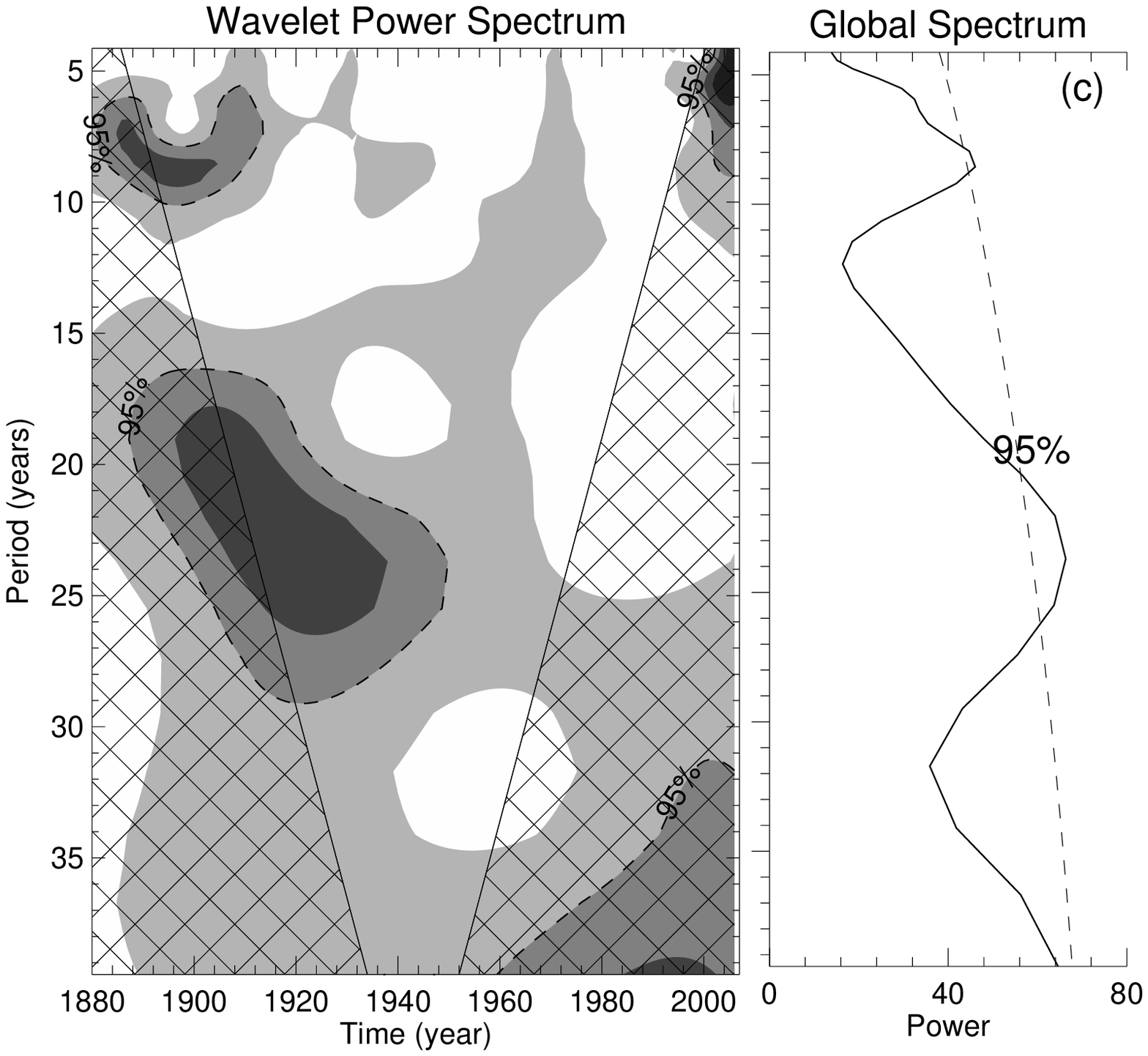}}
\caption{The same as Fig.~11, but for the variations 
in the mean meridional motion of spot groups in the different 
$10^\circ$ latitude intervals, 
(a) $0^\circ$\,--\,$10^\circ$,  
 (b) $10^\circ$\,--\,$20^\circ$, and (c) $20^\circ$\,--\,$30^\circ$,
of the northern (upper panel) 
and the southern (lower panel) hemispheres.} 
\end{figure*}

\section{Conclusions and discussion}
From the analysis of the largest available reliable  sunspot group data;  
 $i.e.$, the combined Greenwich and SOON sunspot group 
data during the period 1879\,--\,2008, we find the following.  
\begin{enumerate}
\item The mean meridional motion  of the sunspot groups varies 
considerably 
on  5\,--\,20 year timescales during the period 1879\,--\,2008.
 The maximum amplitude of the variation is 10\,--\,15 m s$^{-1}$.
\item The pattern  and amplitude  of the solar cycle 
 variation in the mean motion are 
  significantly different during the different cycles.
 During the maximum epoch (year 2000) of the current cycle, the mean  motion
is relatively stronger than in the past  $\approx$100 years, and it is 
northbound   in both the
 northern and
the southern  hemispheres.
\item The north-south difference (north-south asymmetry) in the 
mean meridional motion of the spot groups also varies with a maximum 
amplitude  of about 10 m s$^{-1}$. The north-south difference was 
  considerably  larger during the early cycles 
(with a strong contribution from the 
high latitudes). It is
   negligible 
 during the recent cycles.
\item Power spectral analyses suggest that  $\approx$ 
3.2- and $\approx$ 4.3-year 
 periodicities exist in the mean meridional motion of the spot groups of 
  the southern hemisphere, whereas  
a  13\,--\,16 year  periodicity is  found 
 in the mean motion of  
 the spot groups of  the northern hemisphere. 
\item The $\approx$ 12- and $\approx$ 22-year periodicities 
 are found to exist 
 in the north-south difference of the mean  motion.
 \item There is a considerable latitude-time dependence in the 
  periodicities of the mean meridional motion of the spot groups. 
There is a strong  
 suggestion that,  in the $10^\circ$\,--\,$20^\circ$  latitude-interval 
of the northern hemisphere,    
a periodicity slowly evolved from $\approx$ 16 year to $\approx$ 10 year,
 over the period 1880\,--\,2007,  
and it  
evolved in the  opposite way,  $\approx$ 10 year to $\approx$ 16 year,  in  
 $20^\circ$\,--\,$30^\circ$ latitude interval.
\end{enumerate}

The behavior of the mean motion of the spot groups in cycle~23 is similar to 
that of  cycle~14, which is a low-amplitude (lowest in 
the last century) and considerably long-duration cycle.   
Cycle~23 is also a relatively  low-amplitude and long-duration cycle, so that 
 the result above (conclusion (2)) may be a part of a real 
long-term behavior in the mean 
meridional motion of the spot groups; $i.e.$, most probably it is 
not an 
 artifact of the  differences (if any)  
between the  Greenwich and the SOON datasets,  within the continuous 
time series of the combined dataset used here.

 Most of the helioseismic measurements of 
the  meridional flows during the current sunspot cycle~23
 suggest  an increase in the 
  amplitudes of 
the surface and the  subsurface poleward  
 meridional flows   with a decrease in magnetic activity 
 (Gonz\'alez Hern\'andez et al. 2008), whereas we find a strong  northbound 
mean meridional motion of the spot groups  during 
 the maximum of cycle~23.
 A reason for this discrepancy 
  may be that
sunspot motions  may not  represent 
the Sun's plasma motions (D'Silva \& Howard 1994), or 
the motions of the   
 magnetic structures  may
 represent the motions of the deeper 
 layers of the 
Sun's convection zone where these structures are anchored
(Javaraiah \& Gokhale 1997a;  Hiremath 2002; 
Sivaraman et al. 2003; Meunier 2005). In addition, 
 the mean meridional motion  of the sunspot groups may  only represent 
the mean solar meridional plasma motion   
 at  low and middle latitudes, because sunspots data are confined 
to  only these latitudes. 
 The magnetic structures of the only large spot groups 
during their initial days 
 might be anchored near the base of the convection zone 
(Javaraiah \& Gokhale 1997a;  Hiremath 2002; 
Sivaraman et al. 2003), 
 hence might have  
largely equatorward motions (Javaraiah 1999).
 While rising through 
the convection zone, the magnetic structures of  the large spot groups 
may be fragmented into the smaller structures 
(Javaraiah 2003; Sch\"ussler \& Rempel 2005).  
The small structures may move  mainly toward the poles
 (\v{S}vanda et al. 2007). However, as can be seen in Figs.~1\,--\,3
 there are also
 equatorward motions 
(may be due to an   effect of the  
reverse meridional flows), mainly near minima of the cycles where 
 a  spot group is relatively   small. 
 
 Meridional flows can transport magnetic flux and    
  cause cancellation/enhancement of magnetic flux, and  
 it is  believed that
poleward meridional 
flows play a major role  in the polarity reversals of the polar magnetic
fields ($e.g.$, Wang 2004). 
The $\approx$ 12-year and the $\approx$ 22-year periodicities of 
the north-south difference in  
the  mean meridional motion of the spot groups 
  may have a close relationship with the 11-year  
solar activity  (the 
emerging  magnetic flux) cycle and the 22-year solar magnetic cycle, 
respectively.  
 Many of the other  periodicities found here  also   
exist in several activity phenomena (Knaack et al. 2005; 
Song et al. 2009, and references therein)
 and solar differential rotation determined from sunspot data 
(Javaraiah \& Gokhale 1995, 1997b; Javaraiah \& Komm 1999; 
Braj\v{s}a et al. 2006). 
   They may be closely 
related to the Rossby type waves that were discussed by Ward (1965) 
and others
(e.g., Knaack et al. 2005; Chowdhury et al. 2009).

  According to the well known Gnevyshev and Ohl rule (G-O rule),   
an odd cycle is stronger than its immediately  preceding even cycle 
(Gnevyshev \& Ohl 1948). Cycle pair~22,23 violated the G-O rule.    
The duration  of the current cycle~23 is very long,  and 
 during the declining phase of this cycle,  the activity in the southern
hemisphere  is considerably stronger  than   in the northern hemisphere.
All these properties of the cycle~23 
could be strongly  related to the large and northbound   mean meridional 
motion of the  spot groups during this cycle. As already mentioned above,  
motions of  magnetic structures such as  sunspots mimic 
the motions of deeper layers of the Sun (see also Javaraiah \& Gokhale 2002). 
Therefore, 
 the magnetic flux cancellation/enhancement 
due to the mean meridional motion of sunspot groups  may take place 
in the subsurface layers of the Sun.  
 By considering the  poleward  meridional
 plasma flows detected by surface  Doppler measurements and  
by helioseismology, 
the deeper
 counter-motion (suggested by the mean meridional motion of spot groups) might 
 amplify the action of the near-surface dynamo in the
 southern hemisphere during cycle~23 for  causing
 stronger magnetic activity on 
this hemisphere.

\begin{acknowledgements}
I am thankful to the anonymous referee for the critical review and very 
useful comments and suggestions. I also thank 
 Dr. L. Bertello
 for useful comments and suggestions. 
 Wavelet software was provided by
 C. Torrence and G. Compo, and  is available at 
 {\tt URL: http//paos.colorado.edu/research/wavelets/}. The MEM FORTRAN code 
was provided to us by Dr. A. V. Raveendran.
\end{acknowledgements}

\end{document}